\documentclass[twocolumn,showpacs,preprintnumbers,amsmath,amssymb]{revtex4}
\usepackage{graphicx}
\usepackage{dcolumn}
\usepackage{bm}
\begin{document}
\title{Short timescale behavior of colliding heavy nuclei at intermediate 
energies}

\author{S. Hudan}
\author{R.T. de Souza} 
\affiliation{
Department of Chemistry and Indiana University Cyclotron Facility \\ 
Indiana University, Bloomington, IN 47405}

\author{A. Ono}
\affiliation{
Department of Physics, Tohoku University, Sendai 980-8578, Japan}

\date{\today}
  
\begin{abstract}
An Antisymmetrized Molecular Dynamics model is used to explore the
collision of $^{114}$Cd projectiles with $^{92}$Mo target nuclei at 
E/A=50 MeV over a broad range in impact parameter. 
The atomic number (Z), velocity, and emission pattern of the
reaction products are examined as a function of the impact parameter and the
cluster recognition time. 
The non-central collisions are found to be essentially binary in character 
resulting in the
formation of an excited projectile-like fragment (PLF$^*$) 
and target-like fragment (TLF$^*$). The decay of these fragments occurs 
on a short timescale, 100$\le$t$\le$300 fm/c. 
The average excitation energy deduced for the PLF$^*$ and TLF$^*$ `saturates
for mid-central collisions, 3.5$\le$b$\le$6 fm, with
its magnitude depending on the cluster recognition time. 
For short cluster recognition times (t=150 fm/c), 
an average excitation energy as high as $\approx$6 MeV is predicted. 
Short timescale emission leads to a loss of initial correlations
and results in features such as an anisotropic
emission pattern of both IMFs and alpha particles emitted from the 
PLF$^*$ and TLF$^*$ in peripheral collisions.

\end{abstract}

\maketitle

\section{Introduction}

Collision of two heavy-ions at intermediate energies can result in the
production of a multi-particle final state \cite{Bowman91, deSouza91, Marie97}. 
These multi-particle final states have been experimentally characterized by 
a wide variety of signals including fragment multiplicity 
\cite{Bowman91, deSouza91}, 
size distributions \cite{Rivet98, Ogilvie91}, 
emission timescales \cite{Kim91, Cornell95, Cornell96, Beaulieu00}, 
scaling behavior \cite{Elliott02, Scharenberg01, Porile89} 
and the attained excitation energy \cite{Cussol93}.
For large fragment multiplicity, within a thermodynamic approach, 
such multi-fragment states have been interpreted as a transition 
of the finite nuclear system from a liquid to a gaseous phase 
\cite{Gross90, Bondorf85, Pochodzalla95, Viola04a}. Recent work has 
investigated the robustness of this conclusion by examining the 
influence  of the surface, through
the density dependence of the entropy, on the stability of the nuclear 
droplet against fragmentation \cite{Toke03, Sobotka04}. All these approaches
however focus on the thermodynamic stability of the system.
In reality, the decaying system is 
formed by the collision dynamics which may not equilibrate 
all degrees of freedom equally \cite{Moretto93, Viola04}.
In order to understand both the formation and decay of excited 
nuclear systems involved in the collision process, 
microscopic approaches have also  
been followed \cite{Bauer87, Aichelin91, Danielewicz91, 
Schnack97, Morawetz00, Wada04}. 
In order to make 
direct comparison with experimental
data such microscopic models typically utilize a 
a two-stage approach.
In the first phase, a dynamical model is used to describe the collision 
dynamics. Clusters produced in this phase are subsequently de-excited by 
a statistical model.
Such a two-stage approach typically views the statistical decay stage 
as decoupled from the dynamical stage that preceded it. 
In the present work we examine the validity of such a de-coupled 
hybrid approach. Specifically, we utilize a microscopic model, 
the Antisymmetrized Molecular Dynamics model,
to investigate how the collision proceeds on short timescales and how the 
reaction characteristics evolve with impact parameter.
In addition, we examine whether initial correlations, 
existing at short times, 
survive the decay stage and how they are manifested in final distributions.

\section{Description of the AMD model}

To describe the dynamical stage of intermediate energy heavy-ion
collisions, we utilize the antisymmetrized molecular dynamics (AMD)
model \cite{ONOa,ONOb,ONOj,ONO-ppnp}. For the present work, we use the
same version of AMD as Ref. \cite{ONOj} which has been used to describe the
multifragmentation reaction of the central $\mathrm{Xe}+\mathrm{Sn}$
collisions at 50 MeV/nucleon.

The description of the dynamics of fragmentation is, in principle, a very complicated
quantum many-body problem.  In the exact solution of the many-body
time-dependent Schr\"odinger equation, the
intermediate and final states should be very complicated states
containing a huge number of reaction channels corresponding to
different fragmentation configurations.  The AMD model respects the
existence of channels, while it neglects some of the interference
among them.  Namely, the total many-body wave function
$|\Psi(t)\rangle$ is approximated by
\begin{equation}
|\Psi(t)\rangle\langle\Psi(t)|\approx
\int
\frac{|\Phi(Z)\rangle\langle\Phi(Z)|}{\langle\Phi(Z)|\Phi(Z)\rangle}
 w(Z,t)dZ,
\label{eq:AMDensemble}
\end{equation}
where each channel wave function $|\Phi(Z)\rangle$ is parametrized by
a set of parameters $Z$, and $w(Z,t)$ is the time-dependent
probability of each channel.

In AMD, we choose the Slater determinant of Gaussian wave packets as
the channel wave function
\begin{equation}
\langle\mathbf{r}_1\ldots\mathbf{r}_A|\Phi(Z)\rangle \propto
\det_{ij} \biggl[ \exp\Bigl\{-\nu(\mathbf{r}_i-\mathbf{Z}_j/\sqrt{\nu})2\Bigr\}
                               \chi_{\alpha_j}(i) \biggr],
\label{eq:AMDWaveFunction}
\end{equation}
where $\chi_{\alpha_i}$ are the spin-isospin states with
$\alpha_i=p\uparrow, p\downarrow, n\uparrow,$ or $n\downarrow$. Thus,
the many-body state $|\Phi(Z)\rangle$ is parametrized by a set of
complex variables $Z\equiv\{{\mathbf{Z}}_i\}_{i=1,\ldots,A}$, where
$A$ is the number of nucleons in the system.  The width parameter,
$\nu=0.16$ $\textrm{fm}^{-2}$, is treated as a constant parameter
common to all the wave packets.  If we ignore the antisymmetrization
effect, the real part of $\mathbf{Z}_i$ corresponds to the position
centroid and the imaginary part corresponds to the momentum centroid.
This choice of channel wave functions is suitable for fragmentation
reactions, where each single particle wave function should be
localized within a fragment.

Instead of directly considering the probability $w(Z,t)$ in Eq.
(\ref{eq:AMDensemble}), we solve a stochastic equation of motion for
the wave packet centroids $Z$, which may be symbolically written as
\begin{equation}
\frac{d}{dt}\mathbf{Z}_i
=\{\mathbf{Z}_i,\mathcal{H}\}_\text{PB}
+\mbox{(NN coll)}
+\Delta\mathbf{Z}_i(t)
+\mu\,(\mathbf{Z}_i,\mathcal{H}').
\end{equation}
The first term $\{\mathbf{Z}_i,\mathcal{H}\}_\text{PB}$ is the
deterministic term derived from the time-dependent variational
principle with an assumed effective interaction.  The Gogny
interaction \cite{GOGNY} is used in the present work.  The second term
represents the effect of the stochastic two-nucleon collision process,
where a parametrization of the energy-dependent in-medium cross
section is adopted. The two-nucleon collision cross-section used
is the same as in Ref. \cite{ONOj} namely,
\begin{equation}
\sigma(E,\rho)=\min\biggl( \sigma_{\text{LM}}(E,\rho),\ \frac{100\
\text{mb}}{1+E/(200\ \text{MeV})}\biggr),
\end{equation}
The collisions are performed with the ``physical
nucleon coordinates'' that take account of the antisymmetrization
effects, and then the Pauli blocking in the final state is
automatically introduced \cite{ONOa,ONOb}.  The third term
$\Delta\mathbf{Z}_i(t)$ is a stochastic fluctuation term that has been
introduced in order to respect the change of the width and shape of
the single particle distribution \cite{ONOh,ONOi,ONOj}.  In other
words, the combination
$\{\mathbf{Z}_i,\mathcal{H}\}_\text{PB}+\Delta\mathbf{Z}_i(t)$
approximately reproduces the prediction by mean field theories (for a
short time period) for the ensemble-averaged single-particle
distribution, while each nucleon is localized in phase space for each
channel.  The term $\Delta\mathbf{Z}_i(t)$ is calculated practically
by solving the Vlasov equation (for a short time period) with the same
effective interaction as for the term
$\{\mathbf{Z}_i,\mathcal{H}\}_\text{PB}$.  In the present version of
AMD \cite{ONOj}, the property of the fluctuation
$\Delta\mathbf{Z}_i(t)$ is chosen in such a way that the coherent
single particle motion in the mean field is respected for some time
interval until the nucleon collides another nucleon.  The last term
$\mu\,(\mathbf{Z}_i,\mathcal{H}')$ is a dissipation term related to
the fluctuation term $\Delta\mathbf{Z}_i(t)$.  The dissipation term is
necessary in order to restore the conservation of energy that is
violated by the fluctuation term.  The coefficient $\mu$ is given by
the condition of energy conservation. However, the form of this term
is somehow arbitrary.  We shift the variables $Z$ to the direction of
the gradient of the energy expectation value $\mathcal{H}$ under the
constraints of conserved quantities (the center-of-mass variables and
the total angular momentum) and global one-body quantities (monopole
and quadrupole moments in coordinate and momentum spaces).  A complete
formulation of AMD can be found in Refs. \cite{ONOj,ONO-ppnp}.

The statistical decay of relatively small primary fragments ($Z<20$)
is calculated by using the code \cite{MARUb} based on the sequential
binary decay model by P\"uhlhofer \cite{PUHLHOFER}.  The code employed
in the present work also takes account of the emission of composite
particles not only in their ground states but also in their excited
states with the excitation energy $E^*\le 40$ MeV.  The experimental
information is incorporated for known levels of $A\lesssim28$ nuclei,
while the Fermi-gas level density is assumed otherwise.  For the
statistical decay of large primary fragments ($Z\ge20$), the decay
code GEMINI \cite{Charity01} is employed. 
In considering the decay of the fragments, both the excitation energy 
and decay probabilities are calculated for spherical fragments independent 
of the true shape of the fragments induced by the reaction dynamics.
The effect of n-p asymmetry, excitation energy, and deformation on the 
nuclear level density are not considered in the decay. 
Introduction of a deformation dependence of the nuclear level density, 
and in particular the treatment of the continuum, results in a significant 
modification of the emission rate for fragments that are weakly bound
or at high excitation \cite{Charity05}.

The system we have chosen to study is $^{114}$Cd + $^{92}$Mo at 
E/A = 50 MeV which can be considered representative of symmetric 
heavy-ion collisions in this energy domain. 
We sampled all impact parameters, b, in the
interval 0$\le$b$\le$b$_{max}$ with a triangular distribution. The maximum
impact parameter b$_{max}$ had a value of 12 fm. The 
touching sphere configuration distance, given by 
R=1.2*(A$_P$$^{1/3}$+A$_T$$^{1/3}$), is equal to 11.2 fm.
The projectile and target were therefore placed at an initial distance of 13 fm
for b$\ge$6.5 fm and 9.8 fm for b$<$6.5 fm.
For a given collision, the fate of the colliding system was followed
until 300 fm/c. At regular intervals, the positions and momenta of all 
nucleons in the system were recorded. At a selected time 
(typically 300 fm/c), which we designate the cluster recognition time, 
the nucleon distributions are subjected to a 
cluster recognition algorithm based on the distance between nucleons. 
The nucleons and clusters that result from 
cluster recognition are 
subsequently propagated along Coulomb trajectories and allowed to 
statistically decay. The identity and momenta of the
final reaction products are recorded for subsequent analysis. In order to 
examine the predictions of this model in a statistically 
significant manner, we have amassed $\approx$25,000 collisions. 
The calculations were performed on a 646 CPU parallel processor system 
of which each CPU was either a PowerPC or Power3+. A single collision 
for this reaction
required 12 to 24 CPU-hours on a node depending on the impact parameter.

\section{General Reaction Characteristics}

\begin{figure}
\vspace*{7.0in} 
\includegraphics{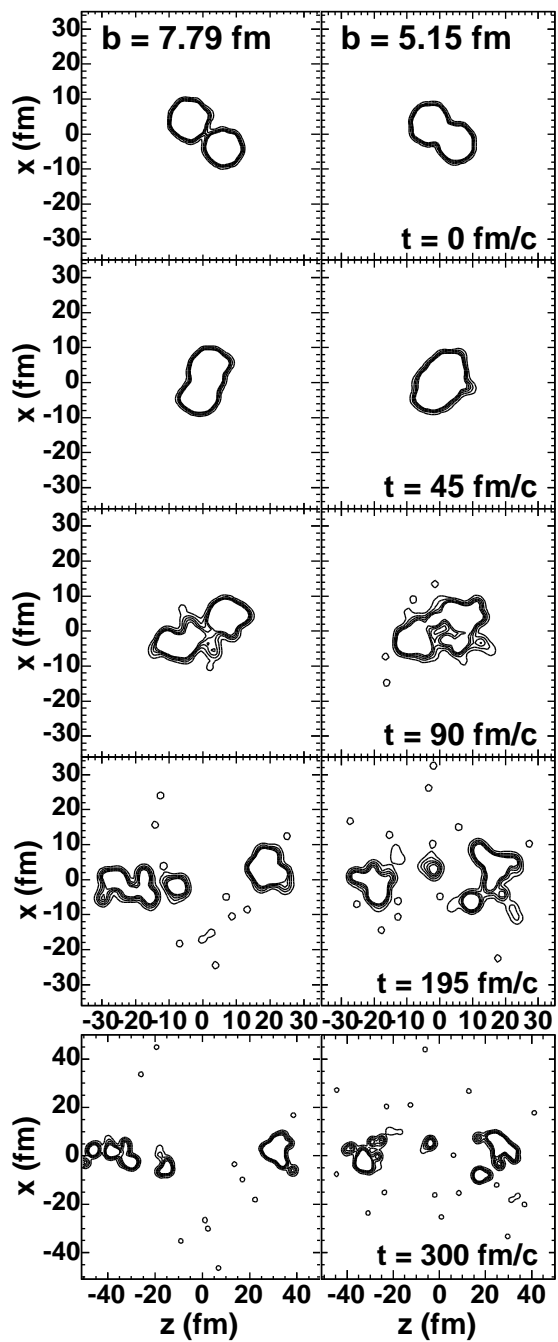}
\caption[]
{Contour diagram depicting the nucleon density distribution in 
spatial coordinates
as a function of time for the reaction $^{114}$Cd + $^{92}$Mo at 
E/A = 50 MeV. The positive z direction corresponds to the direction of 
the projectile. The columns correspond to different 
impact parameters, b=7.79 fm (left) and b=5.15 fm (right).} 
\label{fig:density}
\end{figure}

Depicted in Fig.~\ref{fig:density} is the density distribution 
of nucleons in R-space as a 
function of time for a mid-peripheral (b=7.79 fm) and mid-central (b=5.15) 
collision. The initial moment in time (t=0) is taken as the near touching 
configuration of the projectile-target system previously described, 
with the projectile approaching the target nucleus 
from the negative z direction.
As the di-nuclear system rotates, the initial dumb-bell 
shape of the two touching nuclei shown in the top panel evolves. 
While in contact, the two nuclei exchange mass, charge, and 
energy, governed by nucleon-nucleon scattering within the mean field. 

For the presented event with b=7.79 fm,
one observes that 
two large nuclei emerge from the collision at t=90 fm/c 
revealing the intrinsically binary nature of the collision. 
In this case, at longer times the elongation 
of the target-like fragment (left) leads to its breakup into 
multiple intermediate size nuclei.
In the case of the mid-central collision with b=5.15 fm however, 
the situation is more difficult to discern. At 
t=90 fm/c, it is unclear whether the system is disassembling into two or three 
large pieces.
What is apparent is that as the two nuclei separate
from each other, one observes that the density distributions 
reflect the nuclear
interaction between the projectile and target nuclei through the formation
of transiently deformed nuclei. These non-spherical geometries 
persist up to 300 fm/c for different cluster sizes. 
Moreover, for both events presented clusters seem to emerge on a relatively 
fast timescale,
t$\approx$90 fm/c. This early production of clusters 
indicates that the timescale of the 
shape/density fluctuations 
responsible for cluster formation operate on this timescale. 
It should be noted that a considerable fraction, though not all, 
of this early stage cluster emission
is located between the two large fragments that emerge from the collision.
The evolution
of the density distributions presented in Fig.~\ref{fig:density} 
can also be viewed from the context of semi-classical colliding liquid drops.
Formation of the transiently extended nuclear system by the collision
dynamics involves the generation of a considerable amount of 
``surface'' nuclear material as compared to ``bulk'' nuclear material.
In comparison to the original system comprised of the projectile and target
nuclei, the multi-fragment final state with multiple
clusters requires the formation of a significant amount of
surface -- an energetically unfavorable change. 
Thus, once the surface-to-volume ratio has been 
increased by the collision dynamics, the energy cost of the 
system re-organizing
to the multi-fragment final state is considerably reduced. 

We examine the  characteristics of the system immediately following this 
dynamical stage of the collision. The products of the reaction 
at this stage are designated the ``primary'' products
which statistically de-excite to form the final reaction products which we also
refer to as the ``secondary'' products. 
For a large ensemble of events we examine the
evolution of both primary and secondary 
distributions with impact parameter, velocity dissipation,
and cluster recognition time.

\begin{figure}[ht]
\vspace*{3.5in} 
\includegraphics{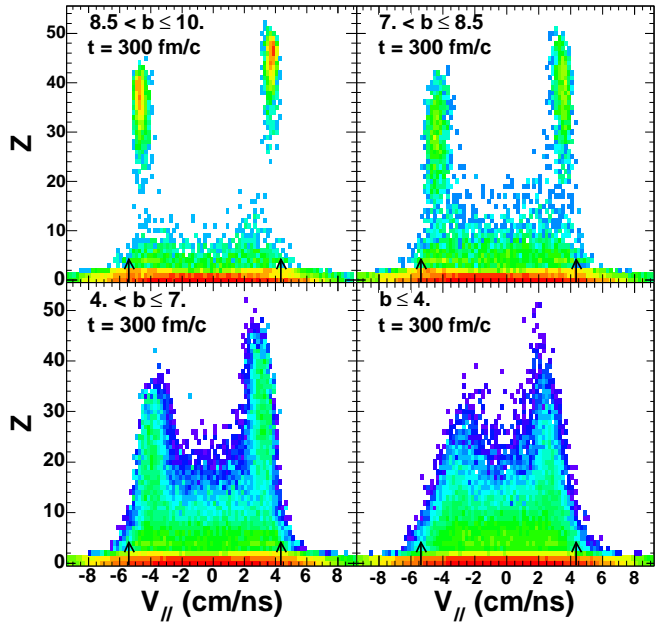}
\caption[]
{(Color online)
Two dimensional diagram of the correlation between the atomic number 
and parallel velocity of
particles at t=300 fm/c for different impact parameters. The arrows 
correspond to the projectile and target velocities. The color scale 
indicates the yield on a logarithmic scale.} 
\label{fig:Zvel}
\end{figure}

An overview of the collisions studied is presented in Fig.~\ref{fig:Zvel}, 
where the correlation between the atomic number  
and parallel velocity (in the center-of-mass frame) 
of particles at t=300 fm/c is examined.
For the most peripheral collisions (8.5$<$b$\le$10 fm) two peaks located at 
Z$\approx$47 and $\approx$39 are clearly evident. 
These peaks correspond to the excited projectile-like (PLF$^*$) 
and target-like (TLF$^*$) nuclei respectively 
and are relatively narrow distributions in velocity 
centered at V$_\parallel$=3.8 and -4.7 cm/ns. Also 
evident is copious production of neutrons (Z=0), hydrogen, 
and helium nuclei. Smaller in yield, are clusters
with Z$\ge$3 and atomic number less than that of the PLF$^*$ and TLF$^*$. 
This pattern, dominated by the survival of the PLF$^*$ and TLF$^*$ for a
peripheral collision, reflects a primarily binary nature. For mid-peripheral 
and mid-central collisions, a similar pattern is observed 
indicating that in this impact parameter range as well
a PLF$^*$ and TLF$^*$ survive the dynamical phase, hence these 
impact parameters are also essentially binary in character. 
With increasing centrality 
$\langle$V$_{PLF^*}$$\rangle$  decreases and 
$\langle$V$_{TLF^*}$$\rangle$ increases reflecting 
an increase in the velocity damping. At the same time, the 
width of the PLF$^*$ and TLF$^*$ velocity distributions increases
indicating the growth of fluctuations.
In addition, with 
increasing centrality the average atomic number of the PLF$^*$ and 
TLF$^*$ decreases while the yield of clusters with 3$\le$Z$\le$15 increases. 
For simplicity, 
we designate the highest Z 
cluster with a velocity larger (smaller) than the center-of-mass 
velocity as the PLF$^*$ (TLF$^*$).
For b$\le$4 fm the decrease in the average Z of the PLF$^*$ combined with the
width of the distribution, lead to an operational definition of 
intermediate mass fragment, namely IMF: 3$\le$Z$\le$10.
Particles with Z$\le$10, manifest broad velocity distributions 
for the most central collisions. 
Examination of the most peripheral collisions reveals a
clear pattern of how the velocity distribution evolves 
with the atomic number (Z) of the fragment.
Neutrons and hydrogen nuclei in particular  have velocity distributions 
that are centered on velocities between those of the PLF$^*$ and
TLF$^*$. In contrast, for nuclei with 3$\le$Z$\le$15
the velocity distribution while broad, 
clearly has a bimodal nature with each of the two peaks centered close to the
PLF$^*$ and TLF$^*$ velocities. This bimodal character is also observed for 
helium nuclei although the distributions are broader. 
These overall patterns manifested for the most 
peripheral collisions are also observed for more central collisions. 

\begin{figure}
\vspace*{3.5in} 
\includegraphics{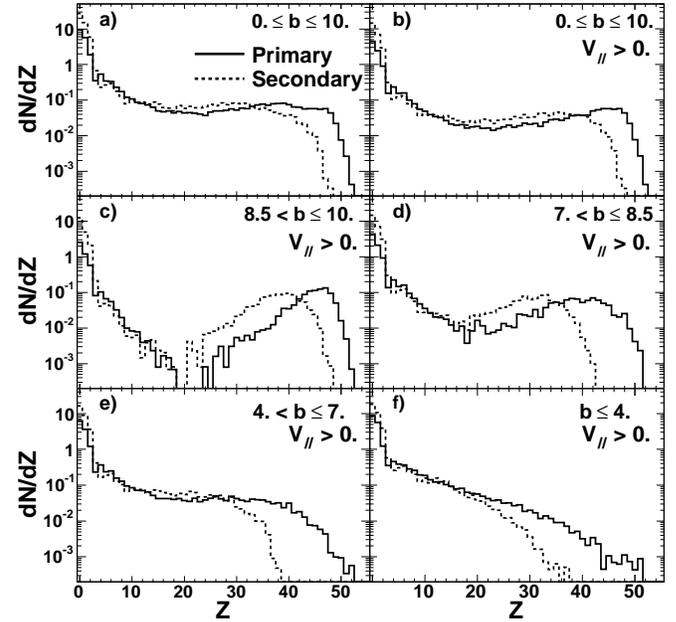}
\caption[]
{Dependence of the primary (solid) and secondary (dashed) Z distributions 
on impact parameter. The differential yield dN/dZ has been normalized by the 
total number of events for each impact parameter interval.} 
\label{fig:Z_dis}
\end{figure}

Depicted in Fig.~\ref{fig:Z_dis} is the dependence of the primary and 
secondary Z distributions on impact parameter. The Z distribution of particles
at t=300 fm/c is the primary distribution and is represented 
as the solid histogram. 
Following Coulomb propagation and statistical decay of the 
excited primary reaction products, the Z distribution of secondary particles 
is represented by the dashed histogram. 
The latter distribution includes both primary fragments that did not 
decay, as well as the decay products of excited primary fragments.
All distributions have been normalized to the total number of events 
for each impact parameter range and therefore 
represent the average multiplicities. 
As may be expected from the trends in Fig.~\ref{fig:Zvel}, 
the charge distribution for the most peripheral collisions, b$>$10 fm,
is largely dominated by two peaks at Z=42 and 48 which correspond to the
TLF$^*$ and PLF$^*$. In Fig.~\ref{fig:Z_dis} we therefore focus on smaller 
impact parameters, b$\le$10 fm. In panel a) the Z distributions 
integrated over impact parameter up to 10 fm are presented.
As expected, the yield for neutrons, hydrogen, and helium is large 
in the primary distribution (solid histogram). 
A large yield is also observed for 3$\le$Z$\le$10.
Evident for Z$\ge$30 is a slight double peak in the primary distribution 
attributable to the presence of the PLF$^*$ and TLF$^*$. 
This double peak structure is eliminated by secondary decay
as it is not evident in the dashed histogram. 
To separate the PLF$^*$ from the  TLF$^*$, 
as well as to crudely separate their decay products, 
we further select particles with the condition V$_\parallel$$>$0.
The resulting primary distribution shown in panel b) 
manifests only a single peak at large Z, which is located at Z=47. 
As observed in panel a) the yield of the Z distribution 
for 3$\le$Z$\le$30 is similar for both the primary and secondary particles.

We examine the dependence of the
Z distribution on impact parameter for V$_\parallel$$>$0 
in  Fig.~\ref{fig:Z_dis}c-f.
For 8.5$<$b$\le$10 fm, Fig.~\ref{fig:Z_dis}c), the primary Z 
distribution is 'V-shaped', reminiscent of the 'U-shape' observed for 
asymmetric fission. The minimum yield observed near Z$\approx$20 is 
deep in comparison to the yield at lower and higher Z indicating that 
asymmetric splits are strongly preferred over symmetric splits. 
It is striking that the multiplicity for Z=3-6 is 
approximately the same as that of Z$\approx$47 (the PLF$^*$). 
The yield ratio for Z=3-6 over Z=45-47 is 0.31/0.37$\approx$0.84, 
indicating a process or processes resulting in copious production of light IMFs. 
This similarity in the yield of the light IMF and the PLF$^*$ 
can, for example, be understood as the asymmetric binary decay 
of a precursor PLF$^*$.
Such a perspective is supported by experimental observation. 
For peripheral collisions of two heavy-ions 
at intermediate energies, the phenomenon of dynamical fission is well 
characterized \cite{Bocage00,Davin02,Colin03}. This dynamical fission has been 
associated with the deformation of the PLF$^*$ induced by the 
collision process. The defining characteristics of this process are the
aligned asymmetric binary decay of the PLF$^*$ and large relative velocities 
between the two produced fragments. 
On general grounds one expects that this dynamical process should 
depend sensitively on both the induced deformation and the excitation 
of the PLF$^*$ \cite{Piantelli02}.  
It is important to observe that the shape of this primary distribution 
largely survives the process of secondary decay. 
The main difference between the primary and secondary distributions 
is that the high Z peak is shifted to lower Z and increases in width. 
For 7$<$b$\le$8.5 fm, Fig.~\ref{fig:Z_dis}d), 
the shape of the primary distribution is better described as 
a 'U-shape'. In contrast to 
the previous impact parameter interval, the minimum located 
at Z$\approx$20 is shallow. This decrease in the depth of the minimum 
can be associated with the increase in 
the probability of
symmetric binary splits relative 
to asymmetric binary splits. This change of the Z distribution with decreasing
impact parameter can be related to an increase in the 
excitation energy of the PLF$^*$. In this impact parameter interval, 
the yield for 
Z=3-6 is significantly larger than that for Z$\approx$42. The ratio of 
the yield of Z=3-6 over the yield of Z=41-43 is 0.66/0.19$\approx$3.47, 
a change by a factor of $\approx$4 as the impact parameter
decreases from 8.5$<$b$\le$10 fm 
to 7$<$b$\le$8.5 fm. This increase in the ratio is due to both an 
increase in the IMF yield 
by a factor of 2 and a decrease in the yield in the vicinity of the 
PLF$^*$ peak. 
The latter decrease reflects the increasing width of the peak in the Z 
distribution attributable to the PLF$^*$ with decreasing impact 
parameter. 
Following secondary decay the 'U-shape' is somewhat less pronounced. 
For yet more central collisions, a 'U-shape' distribution is not observed even 
for the primary distribution. In panel e) no clear bump is observed at large 
Z, indicating the decreased likelihood that a high Z PLF$^*$ survives to the
cluster recognition time of t=300 fm/c. 
For the most central collisions shown, b$\le$ 4 fm, 
the primary Z distribution is exponential over a large range in Z. 
This exponential behavior of the yield is suppressed for Z$\ge$30 
due to the finite size (atomic number) of the system. 
The secondary Z distribution for central collisions also 
exhibits an exponential character for Z$>$3, 
although the onset of the finite size effects is observed at Z=20.
The main effect of secondary decay on the Z distribution, 
for all impact parameters 
is to significantly
enhance the yield of neutrons, hydrogen, and helium nuclei, while decreasing 
the maximum Z observed, namely the atomic number of the PLF$^*$ and TLF$^*$.

\begin{figure}
\vspace*{4.0in} 
\includegraphics{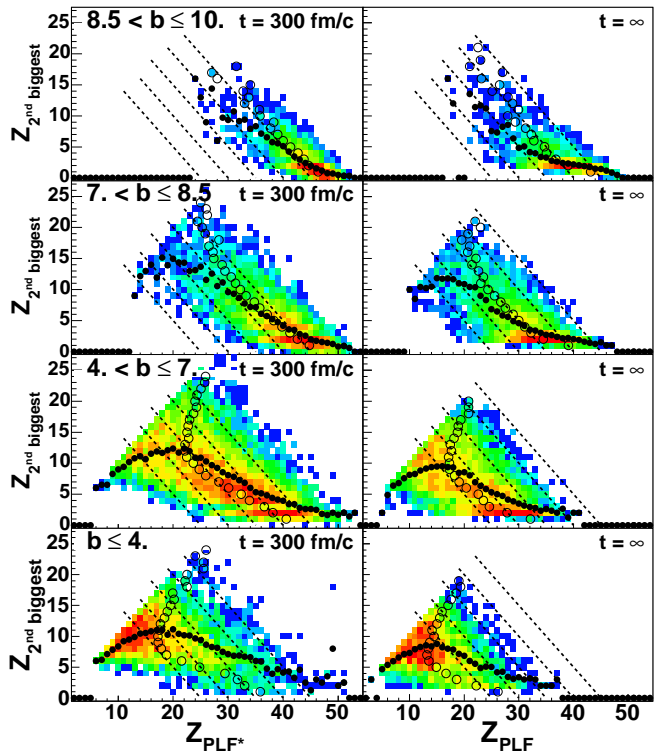}
\caption[]
{(Color online)
Correlation between the atomic number of the fragments with the 
largest and second largest atomic number that have velocities larger 
than the center-of-mass velocity.
The solid circles represent the average charge of the second biggest fragment 
for a given charge of the biggest fragment. 
The open circles represent the average charge of the biggest fragment 
for a given charge of the second biggest fragment.} 
\label{fig:charge_correl}
\end{figure}

We have investigated whether the similarity of the yield for Z=3-6 
and Z$_{PLF^*}$ 
for peripheral collisions is an indication that the two largest fragments
forward of the center-of-mass originate from a common parent. 
Displayed in the
two-dimensional diagrams of Fig.~\ref{fig:charge_correl} 
is the joint probability of observing the 
largest and second largest fragments 
both with V$_\parallel$$>$0. 
For reference the dashed lines correspond to 
Z$_{TOT}$ = Z$_{PLF^*}$ + Z$_{2^{nd} biggest}$ = 45, 40, 35, 30 and 25.
The distribution at t=300 fm/c is presented in the left column while
the distribution following secondary decay is shown in the right column.
In the case of primary fragments (left column), for b$>$4 fm, a clear 
anti-correlation is observed between the atomic number of 
the largest and second largest fragment. 
In order to examine the average behavior of the two dimensional distribution, 
we also indicate as solid and open circles the 
$\langle$Z$_{2^{nd} biggest}$$\rangle$ for a given Z$_{PLF^*}$ and 
$\langle$Z$_{PLF^*}$$\rangle$ for a given Z$_{2^{nd} biggest}$. 
Strong correlation of $\langle$Z$_{PLF^*}$$\rangle$ and 
$\langle$Z$_{2^{nd} biggest}$$\rangle$ is evidenced by the near 
overlap of the open and closed circles over an extended range. 
Divergence of the symbols indicates that either the two fragments 
do not originate from a common parent or that finite size effects 
strongly influence
the observed correlation.
For 8.5$<$b$\le$10 fm, Z$_{TOT}$ is almost constant over the range 
of Z$_{PLF^*}$ with a value of $\approx$ 45. 
This value corresponds to an average loss of three charges from the 
incident Cd nucleus with the observed
anti-correlation signaling a conservation of charge between the
largest and second largest fragment. This anti-correlation signals that 
both fragments do on average originate 
from a common parent fragment. For a second largest fragment with Z=6, the
$\langle$Z$_{PLF^*}$$\rangle$ is $\approx$40 consistent with dynamical 
breakup \cite{Montoya94,Davin02}.
While mid-peripheral (mid-central) collisions exhibit an anti-correlation,
Z$_{TOT}$ changes from $\approx$45 (45) at high Z$_{PLF^*}$ to
$\approx$40 (35) at low Z$_{PLF^*}$. This 
change in Z$_{TOT}$
might indicate that the PLF$^*$ 
splits into three or more pieces or simply reflect the changing size of the 
parent fragment over the finite impact parameter interval considered.
For b$\le$4 fm, the average atomic number of the largest 
and second largest fragment are closer, 
$\langle$Z$\rangle$=20 and 9 respectively, 
as one might expect for a binary decay at high excitation, consistent
with the increased probability of symmetric splits indicated by the Z 
distribution in Fig.~\ref{fig:Z_dis}. The 
general trends observed for the primary fragments are also evident 
following decay in the
charge correlation of secondary fragments (right column). 
Similar charge correlation patterns have been experimentally 
observed \cite{Davin02} indicating a transition from asymmetric
splits toward those in which all asymmetries are populated.
The total charge of the two fragments after secondary decay is typically 
reduced by 
5 to 10 charges as compared to the total charge of the primary fragments.
Following secondary decay, the emission of Z=4-6 results in 
a marked horizontal line in the charge correlation. 
This feature in the charge correlation has also been experimentally observed 
and has been previously attributed to dynamical fission \cite{Davin02}.

\section{Velocity Dissipation of the PLF$^*$ and its Excitation}

\begin{figure}
\vspace*{3.5in} 
\includegraphics{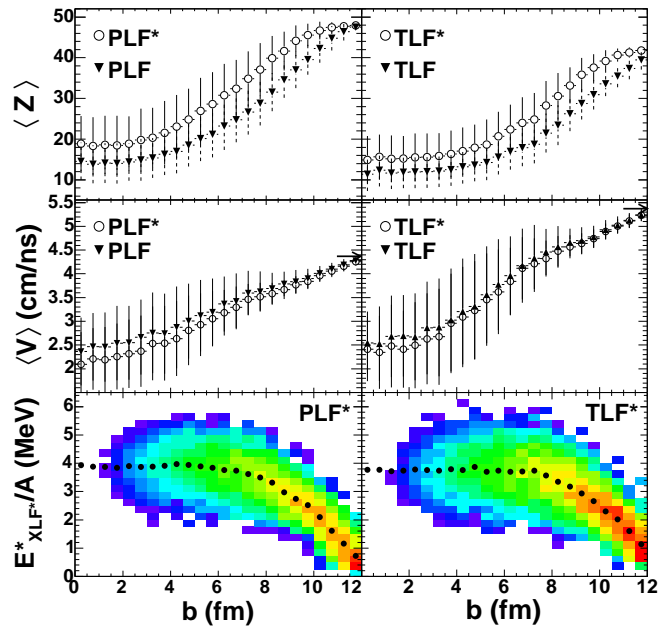}
\caption[]
{(Color online)
Dependence of the $\langle$Z$\rangle$,$\langle$V$\rangle$, and E$^*$/A of the 
PLF$^*$ and TLF$^*$ on  impact parameter. 
Error bars indicate the standard 
deviation of the distribution. In the bottom panels the solid symbols denote
the $\langle$(E$^*$/A)$_{XLF^*}$$\rangle$ as a function of b.} 
\label{fig:Z_V_Estar}
\end{figure}

A more quantitative picture of the evolution of the general properties of the 
PLF$^*$ and TLF$^*$ with impact parameter 
is displayed in Fig.~\ref{fig:Z_V_Estar}. For the most 
peripheral collisions studied, b$>$10 fm, 
the $\langle$Z$_{PLF^*}$$\rangle$ is $\approx$48, the atomic number of the 
projectile. The $\langle$Z$_{PLF^*}$$\rangle$ 
decreases smoothly with decreasing impact parameter until 
b$\approx$3-4 fm.  For smaller 
impact parameters, $\langle$Z$_{PLF^*}$$\rangle$ shows no dependence 
on impact parameter and has a value of $\approx$19. 
For b$<$10 fm, $\langle$Z$_{PLF}$$\rangle$, 
namely the average atomic number following decay, 
is approximately 4-9 units less than $\langle$Z$_{PLF^*}$$\rangle$ and 
exhibits the same impact parameter dependence as $\langle$Z$_{PLF^*}$$\rangle$.
It should be noted that the largest difference 
between $\langle$Z$_{PLF^*}$$\rangle$ and 
$\langle$Z$_{PLF}$$\rangle$ is observed for mid-peripheral collisions 
with an impact parameter $\approx$8 fm. 
The average center-of-mass velocity of the PLF$^*$,
$\langle$V$_{PLF^*}$$\rangle$, also exhibits a smooth 
dependence on impact parameter, decreasing monotonically 
from $\langle$V$_{PLF^*}$$\rangle$ $\approx$ 4.3 cm/ns 
for the most peripheral collisions to $\approx$2.5 cm/ns for b=3 fm. 
For more central 
collisions $\langle$V$_{PLF^*}$$\rangle$ only shows a weak 
dependence on impact parameter. 
With increasing centrality the width of the velocity damping distribution 
(indicated by the error bars) increases significantly, 
indicating the growth of fluctuations.

\begin{figure}
\vspace*{4.0in} 
\includegraphics{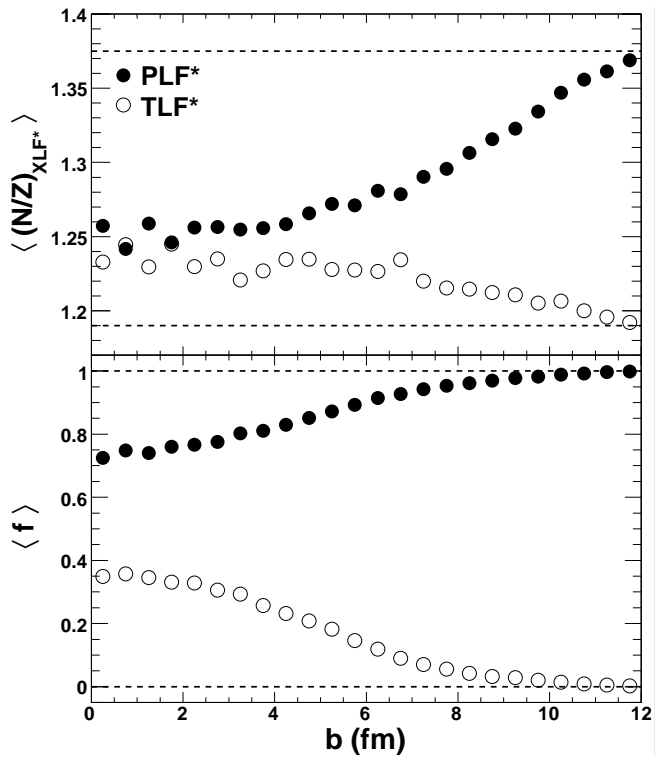}
\caption[]
{Top panel: Dependence of the $\langle$N/Z$\rangle$ of 
the PLF$^*$ and TLF$^*$ on impact parameter.Dotted lines indicate the 
initial N/Z of projectile and target nuclei.
Bottom panel: Fraction, f, of nucleons found in the PLF$^*$ (solid symbols) 
or TLF$^*$ (open symbols) that originate from the projectile.} 
\label{fig:NZ}
\end{figure}

The predicted velocity damping 
of the PLF$^*$ evident in the middle panel is associated 
with a corresponding increase in the excitation of the PLF$^*$ 
as shown in the bottom panel of Fig.~\ref{fig:Z_V_Estar}. 
Such an association between velocity damping and excitation 
has been experimentally observed \cite{Yanez03}.
While the average E$^*$/A of the PLF$^*$ rapidly increases 
with impact parameter for peripheral collisions, it saturates 
at $\approx$4 MeV by b=6 fm. 
The trends observed for the PLF$^*$ are also observed for the TLF$^*$ as 
depicted in the right column of Fig.~\ref{fig:Z_V_Estar}.
It is interesting to note that the $\langle$E$^*$/A$\rangle$ 
for small impact parameters attained for both the 
PLF$^*$ and TLF$^*$ is the same despite the smaller size
of the TLF$^*$ (Z$\approx$15) as compared to the PLF$^*$ (Z$\approx$19). 
This difference of $\approx$20-25\% in Z corresponds to a similar
difference in A (see Fig.~\ref{fig:NZ}). Equal partition of E$^*$ would thus 
result in a larger $\langle$E$^*$/A$\rangle$ for the TLF$^*$ as compared to the
PLF$^*$. An $\langle$E$^*$/A$\rangle$=4 MeV for the PLF$^*$ 
would correspond to an $\langle$E$^*$/A$\rangle$=5 MeV for the TLF$^*$.
The similarity of $\langle$E$^*$/A$\rangle$ for both the PLF$^*$ and TLF$^*$ is
indicative that the degree to which thermalization is achieved is large. 
For the most peripheral collisions, b$\approx$12 fm, 
the non-zero value of the
$\langle$(E$^*$/A)$_{TLF^*}$$\rangle$ and 
$\langle$(E$^*$/A)$_{PLF^*}$$\rangle$ is due in part to 
the mismatch between the binding energy of the projectile and target in AMD
and their real binding energies. This error typically ranges from 0.2 to 0.3 MeV. 
Additional excitation may occur due to the mean field or Coulomb interaction.

The composition of the excited PLF$^*$ and TLF$^*$ that subsequently undergoes
decay
is indicated in Fig.~\ref{fig:NZ}. In this figure the $\langle$N/Z$\rangle$ 
of both the PLF$^*$ and
TLF$^*$ are examined as a function of impact parameter. For b$>$6 fm, the 
$\langle$N/Z$\rangle$ 
of both PLF$^*$ and TLF$^*$ evolves essentially linearly with impact parameter
from the initial values of 1.375 and 1.19 
for the projectile and target respectively.
Over this range of impact parameter, this behavior could be interpreted
as equilibration of N/Z. 
However, the change in $\langle$N/Z$\rangle$ is larger for the
PLF$^*$ as compared to the TLF$^*$ by a factor of two. This difference reflects
the fact that exchange between the PLF$^*$ and TLF$^*$ is not the only process
occurring thus complicating the interpretation of the
change in N/Z in terms of equilibration.
For more central 
collisions, the $\langle$N/Z$\rangle$ remains essentially constant having 
saturated at a value of $\approx$1.24-1.26. The similarity of the average N/Z
value for the PLF$^*$ and TLF$^*$ could be interpreted as equilibration of this 
degree-of-freedom. If this is indeed the case, 
it is interesting to note that for b$\approx$4 fm, this
equilibration is already achieved.
For comparison the 
$\langle$N/Z$\rangle$ of the system is $\approx$1.29. 
The slightly lower N/Z asymptotic value for central collisions 
as compared to the N/Z of the system suggests either 
a preferential emission of free neutrons or the production of neutron-rich 
fragments in the dynamical stage.

We examine the degree to which mixing occurs in the lower panel of 
Fig.~\ref{fig:NZ}. In this figure the dependence of f, the fraction of nucleons
in the PLF$^*$ or TLF$^*$ that originate from the projectile, on impact parameter
is presented. It is interesting to note that for b$\ge$6 fm, the region in
which $\langle$N/Z$\rangle$ changed linearly with b, the fraction of nucleons
in the PLF$^*$ that were originally in the projectile is large, f$\ge$0.9. 
Only for smaller impact parameters does the degree of mixing of projectile and 
target nucleons become larger. Thus, the 
large change in $\langle$N/Z$\rangle$ 
does not require large mixing of the projectile and target nucleons.
It is instructive to note that the quantity f, appears to saturate for b$\le$2 fm 
with a maximum of $\approx$35\% of the PLF$^*$ nucleons originating from the
target. For the TLF$^*$, in the case of small impact parameters, the degree of
mixing is similar.
It has been experimentally demonstrated that 
for mid-peripheral collisions 
the N/Z degree of freedom 
does not reach equilibrium 
\cite{Tsang04}. However, 
the present result indicate that N/Z equilibrium is attained for 
mid-central collisions, despite the incomplete mixing of the projectile and
target nucleons. This result is of significance to future work with radioactive
beams, indicating the degree to which the N/Z exotic projectile can be excited 
while only modestly pertubing its N/Z.

\begin{figure}
\vspace*{3.0in}
\includegraphics{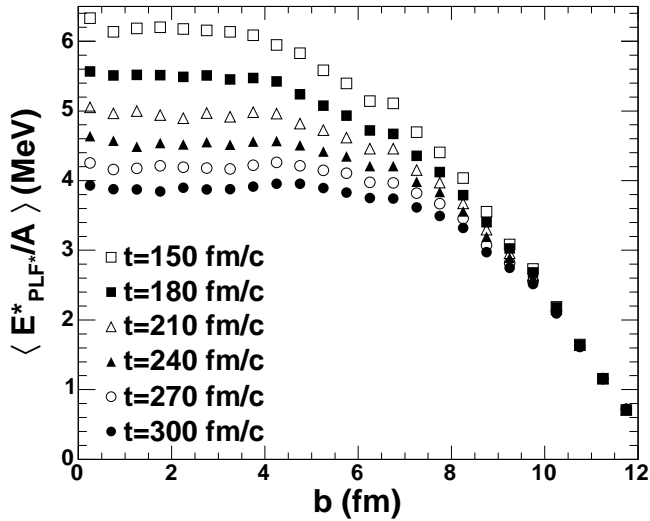}
\caption[]
{Dependence of the $\langle$E$^*$/A$\rangle$ of the
PLF$^*$ on  impact parameter for different cluster recognition times.}
\label{fig:Estar_Time}
\end{figure}

To probe the origin of the saturation in E$^*$/A of the PLF$^*$ and TLF$^*$
for b$<$6 fm observed in Fig.~\ref{fig:Z_V_Estar}, 
we have investigated the influence of our choice of 
cluster recognition time on the excitation energy of the PLF$^*$. 
We have chosen to recognize the clusters 
at t= 150, 180, 210, 240, 270, and 300 fm/c
and compare the dependence of excitation energy on impact 
parameter for the different cluster recognition times. As evident 
in Fig.~\ref{fig:Estar_Time}, while for peripheral collisions 
the average excitation energy is fairly independent 
of the choice of cluster recognition time, with decreasing impact parameter 
the average excitation energy deduced depends significantly on the choice of 
cluster recognition time. 
For different cluster recognition times one also observes that 
the onset of the saturation in excitation energy 
occurs at different impact parameter. For t=300 fm/c the onset of the 
saturation occurs at b$\approx$6 fm ($\approx$25 \% of the cross-section) 
while for t=150 fm/c, 
the onset occurs at b$\approx$4 fm ($\approx$10 \% of the cross-section).
The events associated with the highest excitation attainable
therefore correspond to a 
significant fraction of the
cross-section. 
For central collisions, the excitation 
attained is higher the earlier one recognizes the clusters. For an early
cluster recognition time, t=150 fm/c, a maximum value 
of $\langle$E$^*$/A$\rangle$$\approx$6 MeV is attained in comparison
to $\langle$E$^*$/A$\rangle$$\approx$4 MeV for t=300 fm/c.
Both the trend and magnitude of  $\langle$E$^*$/A$\rangle$
is consistent with the AMD calculations for a more asymmetric system
\cite{Wada04}.
This rapid decrease in $\langle$E$^*$/A$\rangle$ is indicative of 
rapid cooling of 
the PLF$^*$.   
As one may imagine, the choice of a cluster recognition
time less than 150 fm/c becomes increasingly problematic due to  
both the conceptual, as well as practical, problem of  
distinguishing clusters during the 
high density phase of the collision.

\begin{figure*}[ht]
\vspace*{6.2in}
\includegraphics{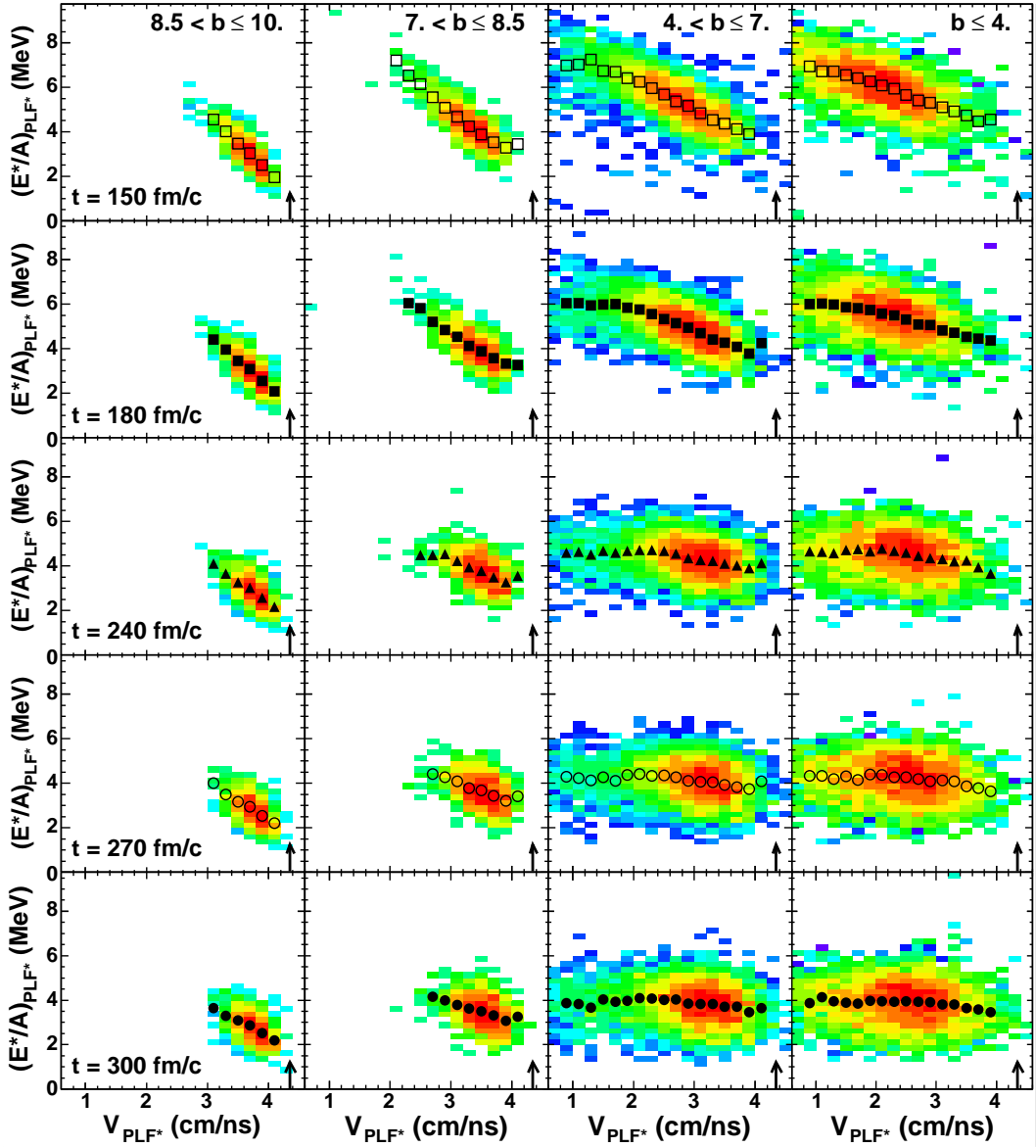}
\caption[]
{(Color online)
Correlation between the E$^*$/A of the
PLF$^*$  and its velocity for different impact parameters and 
cluster recognition times. The symbols indicate the 
average E$^*$/A as a function of V$_{PLF^*}$. The arrows correspond to the 
beam velocity.}
\label{fig:Estar_b_clus}
\end{figure*}

The dependence of $\langle$V$_{PLF^*}$$\rangle$ and 
$\langle$E$^*$/A$\rangle$ on impact parameter 
suggests a direct correlation between these two quantities. 
The correlation between these two quantities 
as a function of both impact parameter and cluster recognition 
time is examined in Fig.~\ref{fig:Estar_b_clus}. 
For peripheral collisions, 8.5$<$b$\le$10 fm (leftmost column), at 
early cluster recognition times, e.g. t=150 fm/c (uppermost panel), 
a narrow anti-correlated 
distribution is observed, namely there is
a strong dependence of the PLF$^*$'s excitation, (E$^*$/A)$_{PLF^*}$, 
on its velocity, V$_{PLF^*}$. 
To more easily examine the correlation between the two quantities, 
the centroid in E$^*$/A for each bin 
in V$_{PLF^*}$ is indicated by the symbol. The significant slope
of $\langle$(E$^*$/A)$_{PLF^*}$$\rangle$ with respect 
to V$_{PLF^*}$ indicates the strong correlation between 
PLF$^*$ excitation and velocity damping.
With increasing cluster recognition time, 
the strong correlation between V$_{PLF^*}$ and (E$^*$/A)$_{PLF^*}$ 
persists although the width of the distribution increases.

For a fixed cluster recognition time, one observes that with 
decreasing impact parameter, the dependence of 
$\langle$(E$^*$/A)$_{PLF^*}$$\rangle$ (symbols) on velocity 
becomes flatter indicating a weakening dependence on average. 
The two dimensional distributions also become broader with increasing 
centrality indicating the growth of fluctuations that attenuate 
the intrinsic correlation between excitation energy and PLF$^*$ velocity.
Examination of the most central collisions studied (b$\le$4 fm) shows that 
while a modest dependence between E$^*$/A and V$_{PLF^*}$ exists 
at t=150 fm/c, for longer cluster recognition times effectively 
no dependence of the PLF$^*$ excitation energy on its velocity 
is observed. At t=300 fm/c,  $\langle$(E$^*$/A)$_{PLF^*}$$\rangle$
does not exhibit any dependence on V$_{PLF^*}$. 
This attenuation of the correlation between
excitation energy and velocity of the PLF$^*$ with increasing
 cluster recognition time is also observed at 
intermediate impact parameters.

\begin{figure}
\vspace*{4.2in}
\includegraphics{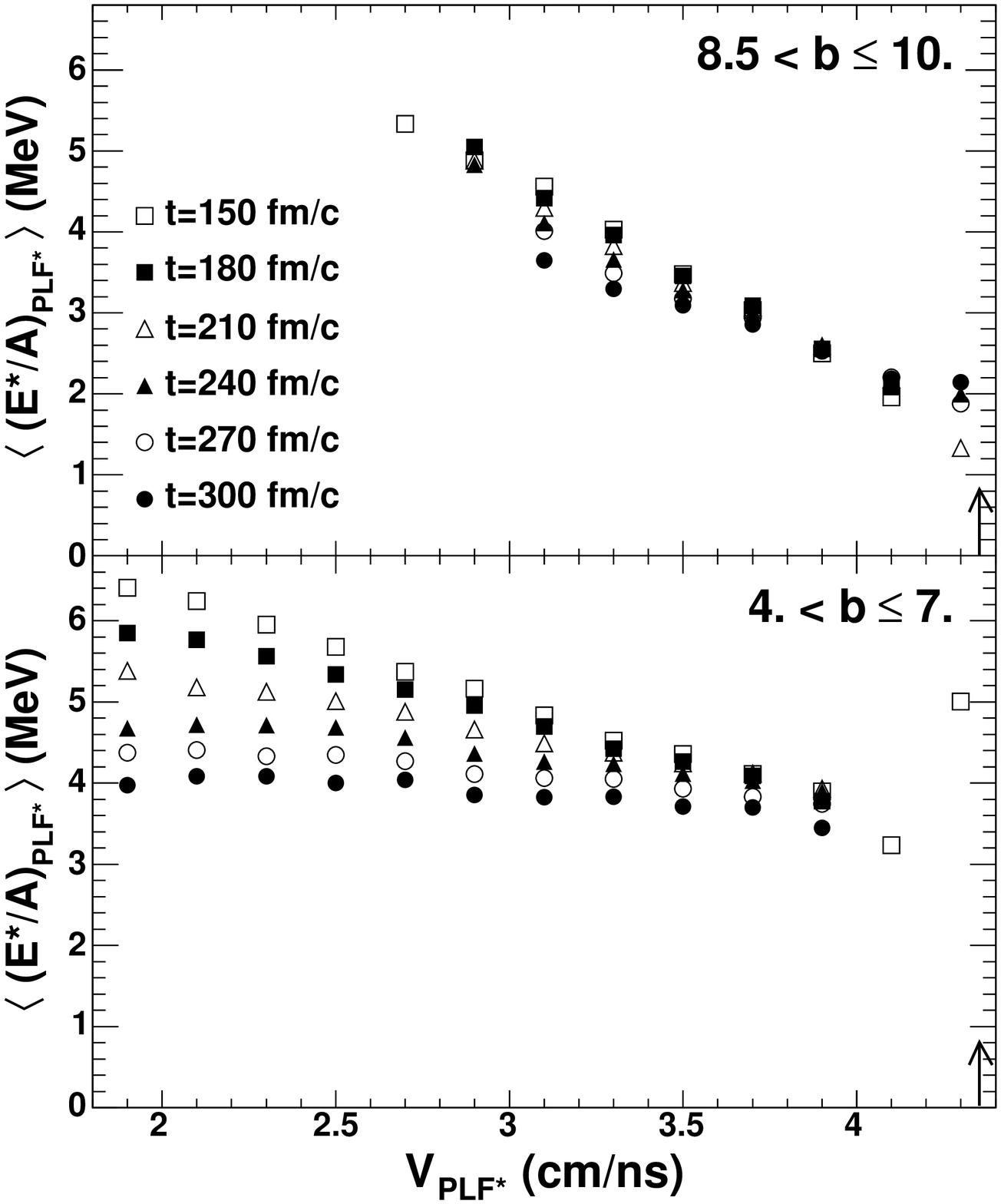}
\caption[]
{Average E$^*$/A of the PLF$^*$  as a function of 
its velocity for different impact parameters and cluster recognition times.}
\label{fig:Estar_Damping_Average}
\end{figure}

To examine the influence of the cluster recognition time 
on the most peripheral collisions in a more quantitative manner, 
we compare in the top panel of 
Fig.~\ref{fig:Estar_Damping_Average} the dependence 
of the average 
excitation energy as a function of V$_{PLF^*}$ for 8.5$<$b$\le$10 fm 
for different cluster recognition times. 
For low to modest velocity damping, i.e. V$_{PLF^*}$$>$3.25 cm/ns, the 
anti-correlation between the average
excitation energy and V$_{PLF^*}$ 
is independent of the cluster recognition time. 
For more damped collisions, however,
one does observe a difference between the calculated average 
excitation energy for different cluster recognition times. 
For V$_{PLF^*}$=4 cm/ns, $\langle$(E$^*$/A)$_{PLF^*}$$\rangle$=2.15 MeV 
while for V$_{PLF^*}$=3.5 cm/ns, 
$\langle$(E$^*$/A)$_{PLF^*}$$\rangle$=3.3 MeV. 
This average excitation of 3.3 MeV is associated with a velocity damping 
from beam velocity of 0.86 cm/ns.

In the lower panel of Fig.~\ref{fig:Estar_Damping_Average} the dependence 
of the average excitation energy as a function 
of V$_{PLF^*}$ for more central collisions 4$<$b$\le$7 fm is shown. 
In contrast to the more peripheral collisions just discussed, 
for all values of velocity damping, the average excitation of the
PLF$^*$ depends on the cluster recognition time.
Even for the smallest velocity damping 
(V$_{PLF^*}$$>$ 3.5 cm/ns)  
a minimum excitation energy of $\approx$3.5 MeV is observed for 
all cluster recognition times. 
Cluster recognition times less than 210 fm/c manifest 
an essentially linear dependence of E$^*$/A on V$_{PLF^*}$ while longer 
cluster recognition times (t$\ge$240 fm/c) exhibit a significantly 
non-linear dependence. By t=300 fm/c,
$\langle$E$^*$/A$\rangle$ is essentially independent of V$_{PLF^*}$.

\begin{figure}
\vspace*{4.0in}
\includegraphics{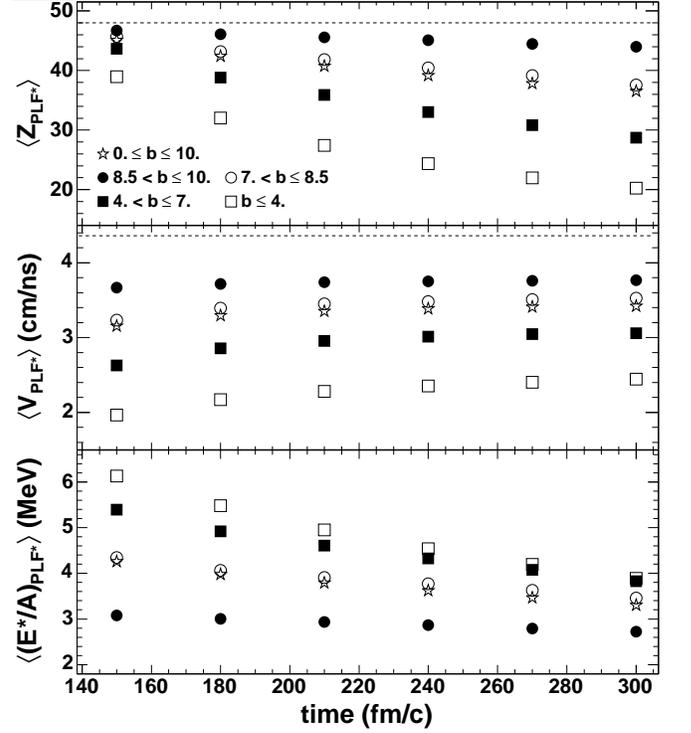}
\caption[]
{Dependence of the $\langle$Z$\rangle$,$\langle$V$\rangle$, 
and $\langle$E$^*$/A$\rangle$ of the PLF$^*$ on cluster recognition time 
for different impact parameters. The dashed line represents the 
projectile atomic number (velocity) in the top (middle) panel.}
\label{fig:ave_prop}
\end{figure}

The dependence of some of the average properties of the PLF$^*$ on both impact
parameter and cluster recognition time are summarized 
in Fig.~\ref{fig:ave_prop}. In the top panel, 
the average atomic number of the PLF$^*$, 
$\langle$Z$_{PLF^*}$$\rangle$, is displayed as a function of 
cluster recognition time for different impact parameters. For the
most peripheral collisions, 8.5$<$b$\le$10 fm, 
and the shortest cluster recognition times,
$\langle$Z$_{PLF^*}$$\rangle$$\approx$47, just below Z$_{BEAM}$=48 
as indicated by the dashed line. Longer cluster recognition times 
result in a slight decrease in $\langle$Z$_{PLF^*}$$\rangle$ to
a value of $\approx$ 44 at t=300 fm/c. This reduction 
in $\langle$Z$_{PLF^*}$$\rangle$ corresponds
to the emission of charge on a short timescale. For more central 
collisions a similar behavior is observed although the magnitude of the 
charge emitted on a short timescale is larger.

In the middle panel of Fig.~\ref{fig:ave_prop}, the trend of 
$\langle$V$_{PLF^*}$$\rangle$ with cluster recognition time 
and impact parameter is presented. For 8.5$<$b$\le$10 fm, 
essentially no change is observed in $\langle$V$_{PLF^*}$$\rangle$ 
as the cluster recognition time changes from t=150 fm/c to 300 fm/c. 
For mid-central and central collisions, a small increase in 
$\langle$V$_{PLF^*}$$\rangle$ is discernible as the 
cluster recognition time increases. 
This slight increase is attributable to the Coulomb re-acceleration
of the PLF$^*$ following the collision combined 
with recoil effects due to predominantly backward emission of particles 
on a short timescale.

The dependence of $\langle$(E$^*$/A)$_{PLF^*}$$\rangle$ 
on cluster recognition time
is depicted in the 
bottom panel of Fig.~\ref{fig:ave_prop} 
for different impact. 
As previously noted in Fig.~\ref{fig:Estar_Time}, for 8.5$<$b$\le$10 fm the 
cluster recognition time has
only a weak influence on $\langle$(E$^*$/A)$_{PLF^*}$$\rangle$. 
Longer cluster recognition
times lead to slightly lower $\langle$(E$^*$/A)$_{PLF^*}$$\rangle$, 
3.1 MeV for t=150 fm/c
as compared to 2.8 MeV at 300 fm/c. More central collisions, however, manifest
a more marked dependence. As apparent in Fig.~\ref{fig:Estar_Time}, 
for b$<$4 fm $\langle$(E$^*$/A)$_{PLF^*}$$\rangle$ reaches 
a value of 6 MeV for the shortest cluster recognition
times, while at longer cluster recognition times 
$\langle$(E$^*$/A)$_{PLF^*}$$\rangle$ is only 
$\approx$4 MeV. This decrease in excitation energy is rapid with most of the 
decrease occurring from t=150-240 fm/c. This rapid decrease in the excitation
energy of the PLF$^*$ is directly related to the emission of particles over
this time interval. As the excited PLF$^*$ rapidly emits charged particles 
between t=150-240 fm/c its atomic number decreases while 
its velocity remains relatively constant. 
Consequently, the correlation between
$\langle$(E$^*$/A)$_{PLF^*}$$\rangle$ and V$_{PLF^*}$ 
observed in Fig.~\ref{fig:Estar_b_clus} is poor 
for central collisions and long cluster recognition times.
Thus, a proper description of this de-excitation of the highly 
excited PLF$^*$ requires modeling the statistical decay of the 
deformed PLF$^*$.

\begin{figure*}
\vspace*{6.2in}
\includegraphics{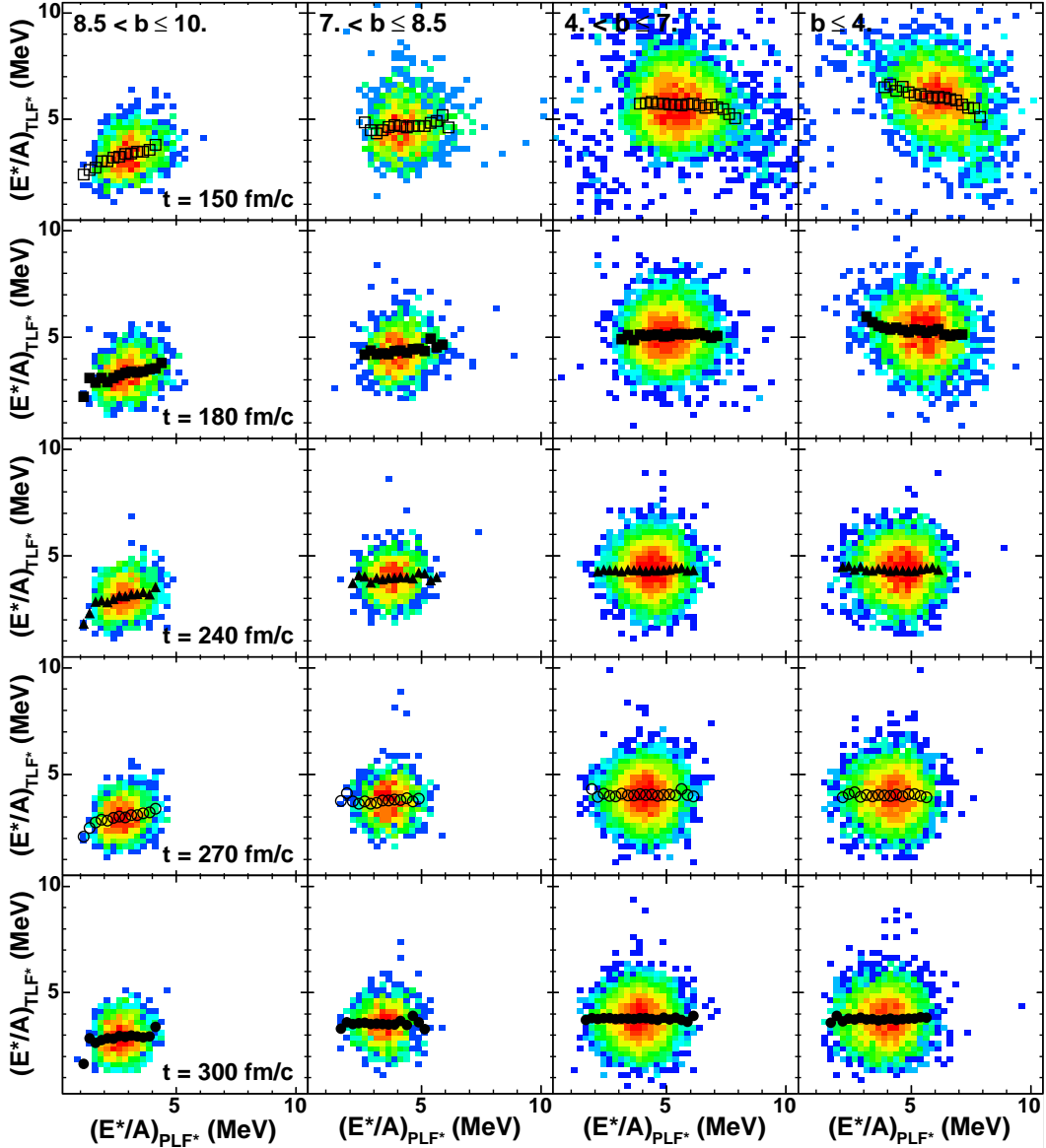}
\caption[]
{(Color online)
Correlation between E$^*$/A of the PLF$^*$ and TLF$^*$
for different impact parameters and cluster recognition times.
The symbols correspond to $\langle$(E$^*$/A)$_{TLF^*}$$\rangle$ 
for a given (E$^*$/A)$_{PLF^*}$.}
\label{fig:Estar_PLF_TLF}
\end{figure*}

An interesting consequence of this rapid emission from the PLF$^*$ 
(and TLF$^*$) is the amelioration of the correlation between the excitation
of the PLF$^*$ and TLF$^*$. Displayed in Fig.~\ref{fig:Estar_PLF_TLF} is the 
two-dimensional distribution of PLF$^*$ and TLF$^*$ excitation energies 
for different impact parameters and cluster recognition times. 
For all impact parameters shown, the distribution is broad 
with the centroid for each (E$^*$/A)$_{PLF^*}$ bin indicated 
by the symbol. For 8.5$<$b$\le$10 fm, a slight positive 
correlation between $\langle$(E$^*$/A)$_{TLF^*}$$\rangle$ and
(E$^*$/A)$_{PLF^*}$ is evident. 
Examination of the correlation between the total excitation, E$^*$, 
of the PLF$^*$ and TLF$^*$ reveals an independence indicating that 
the observed correlation between (E$^*$/A)$_{PLF^*}$ and 
$\langle$(E$^*$/A)$_{TLF^*}$$\rangle$ is principally due to a 
correlation between A$_{PLF^*}$ and A$_{TLF^*}$ for the most peripheral 
collisions.
In contrast, for the most central collisions, b$\le$4 fm, 
an anti-correlation between $\langle$(E$^*$/A)$_{TLF^*}$$\rangle$ 
and (E$^*$/A)$_{PLF^*}$ is evident. 
Both E$^*$ and A of the PLF$^*$ and TLF$^*$ manifest the same anti-correlation.

\section{Emitted Particles}

\begin{figure}
\vspace*{4.0in} 
\includegraphics{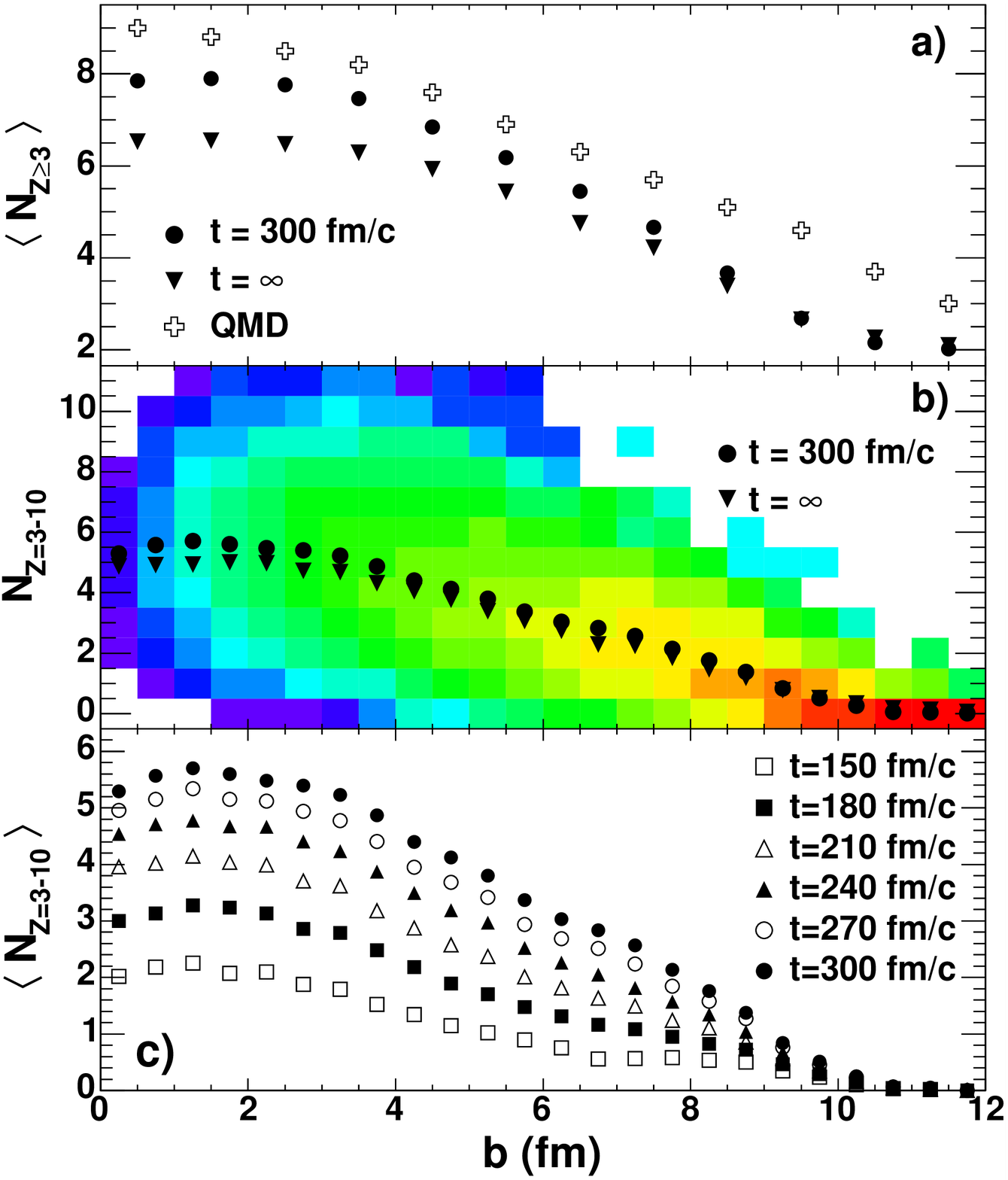}
\caption[]
{(Color online)
Multiplicity of fragments as a function of impact parameter. 
Panel a: Average multiplicity of Z$\ge$ 3 at t=300 fm/c (solid circles), 
after secondary decay (solid triangles) and for QMD (open crosses). 
The QMD results are extracted from \cite{Nebauer99}. 
Panel b: Multiplicity of Z=3-10. 
Panel c: Average multiplicity of Z=3-10 for different 
cluster recognition times.} 
\label{fig:Nimf_b}
\end{figure}

As evident from Fig.~\ref{fig:density}, as the PLF$^*$ and TLF$^*$ separate, 
clusters are produced. 
This fragment production as already demonstrated 
can occur on relatively short time scale impacting the Z, velocity, and
(E$^*$/A) of the PLF$^*$ and TLF$^*$. In order to characterize this 
fast emission process in more detail, 
we examine the multiplicity of fragments 
produced as a function of impact parameter in Fig.~\ref{fig:Nimf_b}. 
Displayed in Fig.~\ref{fig:Nimf_b}a) is the average multiplicity 
of fragments, Z$\ge$3, at t=300 fm/c (solid circles).
One observes that this multiplicity increases 
with decreasing impact parameter and saturates for b$\approx$3 fm. 
For the most peripheral collisions the average multiplicity is 2, 
corresponding to the existence of the only PLF$^*$ and TLF$^*$. 
The average fragment multiplicity reaches a value of 3 at b$\approx$8-9 fm. 
For this impact parameter interval, on average, 
one fragment is produced in coincidence with the PLF$^*$ and TLF$^*$. 
This result is consistent with the asymmetric split of the PLF$^*$ deduced 
from the Z distribution (Figs.~\ref{fig:Z_dis} and \ref{fig:charge_correl}).
For the most central collisions, b$<$3 fm, the average fragments multiplicity 
is constant and is $\approx$8. 
Following secondary decay (solid triangles) 
the fragment multiplicity is reduced slightly 
due to the decay of fragments into particles with Z$\le$2.
For b$>$7 fm the effect of secondary decay on the fragment multiplicity 
is negligible 
while for the most central collisions 
the average multiplicity decreases from 8 to 6.5. 
The increased excitation energy associated with more central collisions is no
doubt responsible for this
increased importance of secondary decay.
The multiplicities predicted in the present calculation are compared to 
those from QMD calculations for the system Xe+Sn \cite{Nebauer99}. 
Although both systems were simulated 
for the same incident energy of 50 MeV/nucleon, the Xe+Sn system is
$\approx$20\% larger in A and $\approx$15\% larger in Z than the present system. 
The multiplicity deduced by QMD (open crosses) is larger that the ones 
of the present work at all impact parameters. 
Given the difference in the system size, the difference between the 
multiplicities for b$\le$6 fm may be reasonable.
The most notable feature of this comparison between the two models 
is the behavior for peripheral collisions, b$\ge$8 fm. The fragment 
multiplicities predicted by AMD appear to be more realistic than those predicted
by QMD. This difference may be due to spurious decay of the
projectile and target in QMD due to the poor description of the ground state 
properties in that model.

The multiplicity distribution of IMFs (3$\le$Z$\le$10) as a function of 
impact parameter is presented in Fig.~\ref{fig:Nimf_b}b) for t=300fm/c. 
While the distribution is narrow for the most peripheral collisions, 
its width rapidly increases with decreasing impact parameter. 
The average IMF multiplicity, indicated by the solid circles, 
evolves from 0 for the most peripheral to $\approx$6 for b=3 fm. 
At an impact parameter of $\approx$9 fm, the average IMF multiplicity 
reaches a value of $\approx$1, consistent with Fig.~\ref{fig:Nimf_b}a). 
The average IMF multiplicity is pretty insensitive to secondary decay as 
indicated by the triangles. 
Comparison between the fragment multiplicity, Fig.~\ref{fig:Nimf_b}a), 
and IMF multiplicity, Fig.~\ref{fig:Nimf_b}b), indicates that even for 
the most central collisions two fragments with a Z$>$10 are present 
at t=300fm/c representing a PLF$^*$ and TLF$^*$ with approximately 20-25 \% 
of the original projectile and target atomic number. 
This result contradicts the physical picture of a single 
source often assumed for central collisions.

Displayed in Fig.~\ref{fig:Nimf_b}c) is the average IMF multiplicity 
dependence on b for different cluster recognition times. 
The average IMF multiplicity increases with 
increasing cluster recognition time for all impact parameters . 
The largest increases are evident for 
the shortest times, t$\le$240 fm/c. 
For all impact parameters, the IMF multiplicity increases by a factor 
of 2 to 3 between t=150 fm/c and t=240 fm/c. After t=240 fm/c, 
the IMF production rate is reduced with an increase of 20-40 \% of 
the IMF multiplicity between t=240 fm/c and t=300 fm/c.

\begin{figure}
\vspace*{4.0in} 
\includegraphics{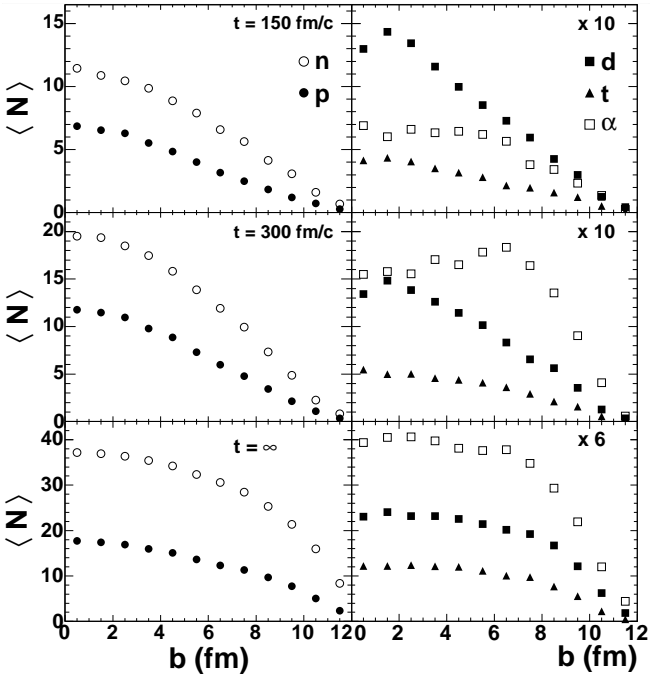}
\caption[]
{Multiplicity of light particles as a function of impact parameter for
different cluster recognition times. 
The multiplicities in the right column have been scaled by the factors 
indicated.} 
\label{fig:Nlp_b}
\end{figure}

The average multiplicity of light charged particles is examined in 
Fig.~\ref{fig:Nlp_b} as a function of both impact parameter and cluster 
recognition time. In the left hand column of Fig.~\ref{fig:Nlp_b} one observes
a monotonic increase of the neutron and proton average 
multiplicities with decreasing
impact parameter both at t=150 fm/c and t= 300 fm/c. At t=150 fm/c a slight 
saturation in the both the neutron and proton multiplicities is observed for 
the most central collision with maximum average multiplicities of 11.5 and 7
attained. A later cluster recognition time of t=300 fm/c results in 
approximately a 50\% increase 
in the multiplicities with the saturation of the multiplicities 
for central collisions being slightly more evident. For this longer
cluster recognition time, the average multiplicities associated with central
collisions are 19.5 and 12 for neutrons and protons respectively. Following 
sequential decay (t=$\infty$), 
one observes a significant increase in the average multiplicities and a 
pronounced saturation in the case of the neutrons. This saturation 
suggests that the total neutron multiplicity, in particular, 
while providing impact parameter selectivity for peripheral collisions is 
a poor selector of more central collisions. Moreover, attempting to 
select central collisions with the neutron multiplicity would on the basis of 
the cross-section be weighted towards mid-central collisions. This result 
explains the experimental observation of the persistence of binary collisions
associated with large neutron multiplicity \cite{Lott92}. Moreover, 
these mid-central 
collisions are associated with the highest average excitation energy attained
as presented in Fig.~\ref{fig:Z_V_Estar}. 

In the case of the light cluster (d,t, and $\alpha$ particle) multiplicities 
a couple of points are noteworthy. For short cluster recognition time 
(t=150 fm/c) the average multiplicity of deuterons is relatively linear 
over the entire impact parameter range and reaches a value of $\approx$1.4
for the most central collisions. Alpha particles, in the case of peripheral
collisions manifest similar multiplicities, however the average multiplicity 
of $\alpha$ particles saturates for b$<$6 fm. Tritons exhibit lower 
multiplicities than both deuterons and alpha particles for all impact 
parameters. For longer cluster recognition time, t=300 fm/c, the deuteron
and triton 
multiplicities remain essentially unchanged as compared to t=150 fm/c. In 
contrast, the $\alpha$ particle multiplicity increases significantly. It is 
interesting to note that the maximum $\alpha$ multiplicity is not associated
with central collisions but rather with b$\approx$7 fm. From this we conclude 
that significant $\alpha$ production/emission, but not 
deuteron or triton 
emission occurs on the timescale 
commensurate with the separation time of the PLF$^*$ and TLF$^*$ 
(150 fm/c$\le$t$\le$300 fm/c). Following 
secondary decay (t=$\infty$) all multiplicities increase significantly. 
Moreover, only for peripheral collisions, b$\ge$8 fm, does the average 
multiplicity of light clusters depend significantly on impact parameter.

\begin{figure*}
\vspace*{4.2in}
\includegraphics{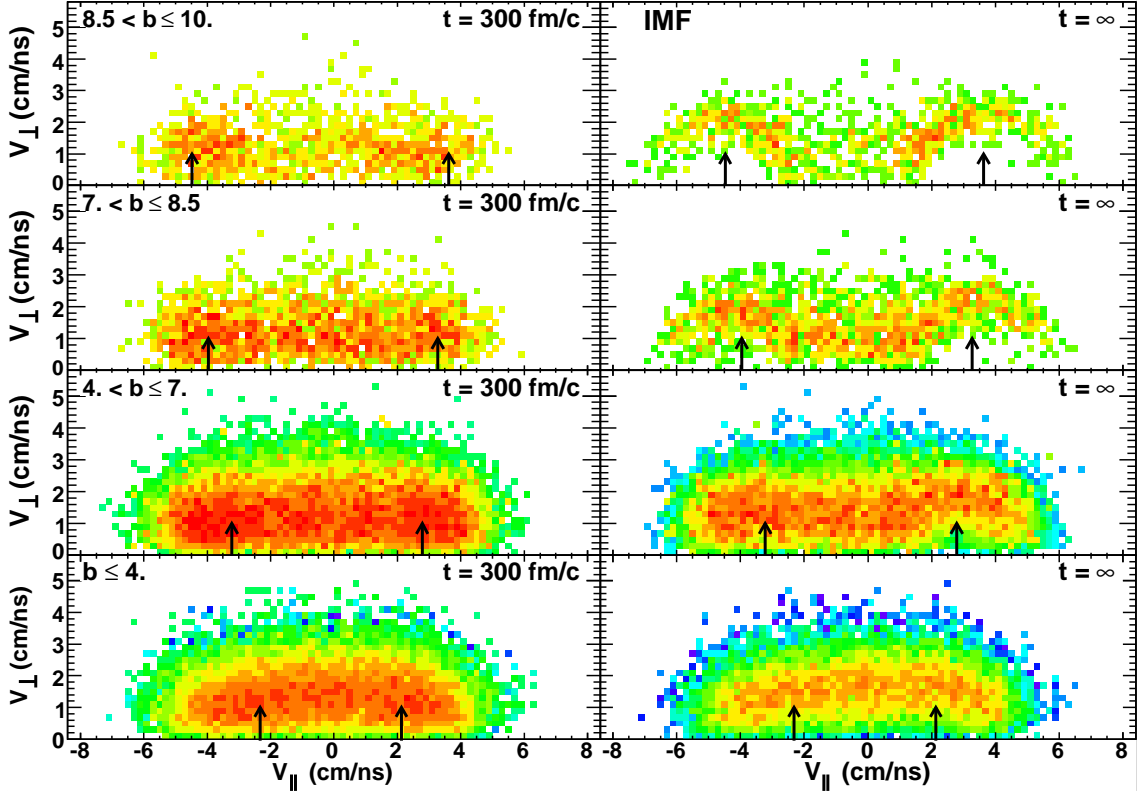}
\caption[]
{(Color online)
Invariant cross-section for IMFs (3$\le$ Z$\le$10) in the COM frame.
The arrows indicate the average parallel velocity of the PLF$^*$ and TLF$^*$.
The color scale indicates the yield on a logarithmic scale. The vertical 
scale in the bottom two
panels has been scaled by a factor of two as compared to the other panels.}
\label{fig:Gal_IMF}
\end{figure*}

We examine the emission pattern for IMFs (3$\le$Z$\le$10) 
in Fig.~\ref{fig:Gal_IMF} both at t=300 fm/c and 
at t=$\infty$ as a function of impact parameter.
In examining the most peripheral collisions for t=300 fm/c, we observe 
two major components which are shifted with 
respect to the velocity of the PLF$^*$ and TLF$^*$ as
represented by the arrows in the figure. In addition a minor component is
visible centered at the velocity of the center-of-mass
i.e. V$_{\parallel}$=0. This emission
pattern is consistent with
anisotropic emission in the frame of 
the PLF$^*$ and TLF$^*$. The most likely origin if the 
observed backward enhancement, 
i.e. towards mid-rapidity, is the asymmetry of the collision process itself.
With increasing centrality, one observes an increase in this 
backward yield, as well as an increase in the yield of the mid-velocity
component. For b$<$7 fm, this mid-velocity yield becomes 
considerable. 
For the most central collisions, the distinct bimodal character 
evident in more peripheral collisions is replaced by a broad distribution.
The impact of Coulomb propagation and secondary decay is shown in the
right column of Fig.~\ref{fig:Gal_IMF}.
In contrast to the broad distributions observed at t=300 fm/c, 
the emission pattern following Coulomb propagation to infinite 
PLF-TLF separation and secondary decay (right column), 
reveals a pattern of two semi-circles centered on the PLF$^*$ and 
TLF$^*$ velocities. Such an emission pattern reflects both the
Coulomb focusing in the field of the separating PLF$^*$ and TLF$^*$, 
as well as emission of
IMFs from the de-exciting PLF$^*$ and TLF$^*$.
For the most peripheral collisions one observes two distinct 
Coulomb circles. 
It is important to note that the intensity pattern
along each of these Coulomb circles is not constant but exhibits a 
significant backward enhancement indicating a memory of the initial 
angular asymmetry. 
With decreasing impact parameter, the center of these 
Coulomb circles shifts toward 
the center-of-mass and increasingly overlap 
as the velocity of the PLF$^*$ and TLF$^*$ decrease. The Coulomb circles also 
become less distinct with increasing centrality reflecting both 
increased excitation of the system and  nucleon-nucleon
scattering.

\begin{figure}
\vspace*{4.0in} 
\includegraphics{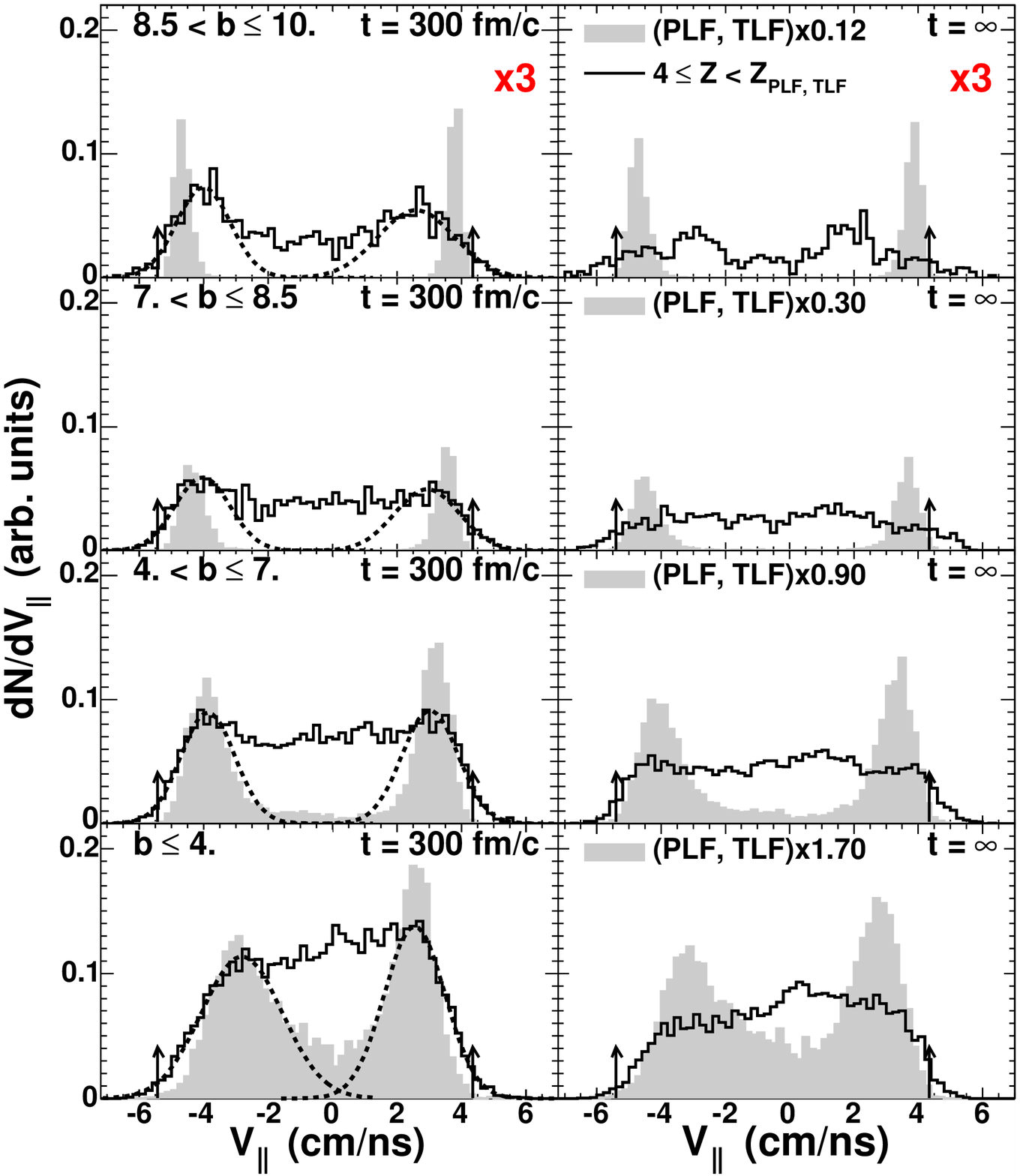}
\caption[]
{Left column: Parallel velocity distributions for the
PLF$^*$ and TLF$^*$ (shaded), 
as well as IMFs (solid histogram) as a function of impact parameter. 
The dashed histograms correspond to a two gaussian fit as described
in the text. 
Right column: Parallel velocity distributions of the PLF, TLF, and IMFs 
following Coulomb propagation and decay.
The IMFs distributions have been normalized to the number of events.
The PLF$^*$, TLF$^*$, PLF and TLF distributions have been
scaled relative to the IMFs distributions.
The arrows indicate the projectile and target velocities.} 
\label{fig:V_dis}
\end{figure}

The parallel velocity distributions of the PLF$^*$, TLF$^*$,
and IMFs and their decay products are shown in Fig.~\ref{fig:V_dis} 
as a function of impact parameter. 
The velocity distributions of the PLF$^*$ and TLF$^*$ (left column) are 
presented for reference (shaded histogram).
For clarity these latter distributions have been scaled relative to 
the IMF distributions by the factors indicated. 
In the case of 8.5$<$b$\le$10 fm, the PLF$^*$ and TLF$^*$ manifest 
gaussian-like velocity distributions that are relatively narrow and 
slightly damped from the beam velocity. With 
decreasing impact parameter, these two distributions move closer in velocity, 
i.e. exhibit increased damping, and become broader. 
The parallel velocity distributions of the PLF and TLF (right column) follow 
the same general trends as those of the PLF$^*$ and TLF$^*$. The widths of the 
secondary large fragments are typically 10-40 \% larger than that of the 
PLF$^*$ and TLF$^*$.

For the most peripheral collisions,
the IMF velocity distribution (solid histogram) is bimodal with 
the most probable values of this two peaked distribution 
displaced toward the center-of-mass velocity as compared to the PLF$^*$ and 
TLF$^*$ velocities, clearly establishing the qualitative trend first observed
in Fig.~\ref{fig:Gal_IMF}.
 In addition to the two gaussian yields attributable to the
emission from the PLF$^*$ and TLF$^*$ an additional IMF component, 
smaller in magnitude, is observed. As previously noted in
Fig.~\ref{fig:Gal_IMF}, 
this additional component has 
an average velocity roughly centered at the center-of-mass velocity. 
For 7$<$b$\le$8.5 fm, the relative magnitude of the mid-velocity 
contribution is increased. 
With increasing centrality, the shape of the IMF velocity 
distribution evolves toward a flat distribution reflecting increased 
fragment production at mid-velocity.

\begin{table}
\caption{\label{tab:Vpar_fit_para}PLF$^*$ and TLF$^*$ average parallel velocity, 
and fit parameters for the two gaussian fit of the IMF 
V$_{\parallel}$ distributions at t=300 fm/c. 
The deduced quantities, $\langle$V$_{\parallel}$$\rangle$ and $\sigma$ 
are expressed in cm/ns.}
\begin{ruledtabular}
\begin{tabular}{ccccccc}
 &\multicolumn{1}{c}{$PLF^*$}&\multicolumn{2}{c}{IMF(PLF$^*$)}&
 \multicolumn{1}{c}{$TLF^*$}&\multicolumn{2}{c}{IMF(TLF$^*$)}\\
 b (fm)&$\langle$V$_{\parallel}$$\rangle$
 &$\langle$V$_{\parallel}$$\rangle$&$\sigma$
 &$\langle$V$_{\parallel}$$\rangle$
 &$\langle$V$_{\parallel}$$\rangle$&$\sigma$\\ \hline
 8.5$<$b$\le$10&3.76&2.59&1.19&-4.65&-4.01&0.84 \\
 7$<$b$\le$8.5&3.51&2.93&1.01&-4.31&-4.03&0.86 \\
 4$<$b$\le$7&2.99&3.02&0.87&-3.57&-3.86&0.84 \\
 b$\le$4&2.26&2.55&0.94&-2.48&-2.81&1.25\\
\end{tabular}
\end{ruledtabular}
\end{table}

We have fit the predicted parallel velocity distributions shown with
two gaussians representing the emission from the PLF$^*$ and TLF$^*$. 
The result is depicted as the dashed histogram in Fig.~\ref{fig:V_dis}. 
The fit parameters for the PLF$^*$ and TLF$^*$ emission are presented in 
Table~\ref{tab:Vpar_fit_para}. 
With increasing centrality
the centroid of the PLF$^*$ velocity distribution decreases and the
centroid of the TLF$^*$ velocity distribution increases as the reaction is
increasingly damped.
While for b$\ge$7 fm, a difference between the average parallel velocity for
IMFs and the PLF$^*$ (or TLF$^*$) is discernible,  
for 4$<$b$\le$7, the IMF distribution is
centered on $\langle$V$_{PLF^*}$$\rangle$. The widths of the distributions
are presented for completeness. No consistent trend of significance
is evident in the extracted widths.

\begin{table}
\caption{\label{tab:Vpar_fit_mult}Average multiplicity of 
4$\le$Z$<$$Z_{PLF^*}$, $Z_{TLF^*}$ for the PLF$^*$, 
TLF$^*$ and mid-velocity (MV) components at t=300 fm/c.
The relative fraction of the mid-velocity component to the total emission
of the PLF$^*$ and TLF$^*$ is also indicated.}
\begin{ruledtabular}
\begin{tabular}{ccccc}
 b (fm)&$PLF^*$&$TLF^*$&MV&P(MV) \\ 
 8.5$<$b$\le$10&0.09&0.09&0.06&0.26 \\
 7$<$b$\le$8.5&0.63&0.63&0.81&0.39 \\
 4$<$b$\le$7&0.99&0.94&1.53&0.44 \\
 b$\le$4&1.61&1.77&1.51&0.31\\
 
\end{tabular}
\end{ruledtabular}
\end{table}

We have also used the two gaussian fits previously described to 
extract the average multiplicity 
associated with the
PLF$^*$, TLF$^*$ and mid-velocity components at t=300 fm/c.
The results are tabulated in Table~\ref{tab:Vpar_fit_mult}. 
With increasing centrality the multiplicities for each component increases
although for the most central collisions, b$\le$ 7 fm the mid-velocity
multiplicity seems to saturate at a value of $\approx$1.5. 
From the peripheral collisions, 8.5$<$b$\le$10, 
to the mid-central collisions, $4<b\le7$, the average 
multiplicity of the PLF$^*$ and TLF$^*$ 
components increases by a factor $\approx$10 with an increase by 
$\approx$25 for the mid-velocity component. 
The relative multiplicity of mid-velocity emission as compared to the total
PLF$^*$ and TLF$^*$ emission increases  
from 0.26 for peripheral collisions to 0.44 for more central 
collisions.

The velocity distributions
of the IMFs are significantly altered by secondary decay.
This influence is most evident for the peripheral collisions 
where the shape of the primary distribution is nearly 
completely destroyed. Naturally, the magnitude of this
secondary decay is particularly sensitive to the excitation 
predicted for the primary fragments. The observed influence of 
secondary decay on the IMF velocity distribution indicates 
that the IMFs are significantly excited.

\begin{figure*}
\vspace*{4.2in}
\includegraphics{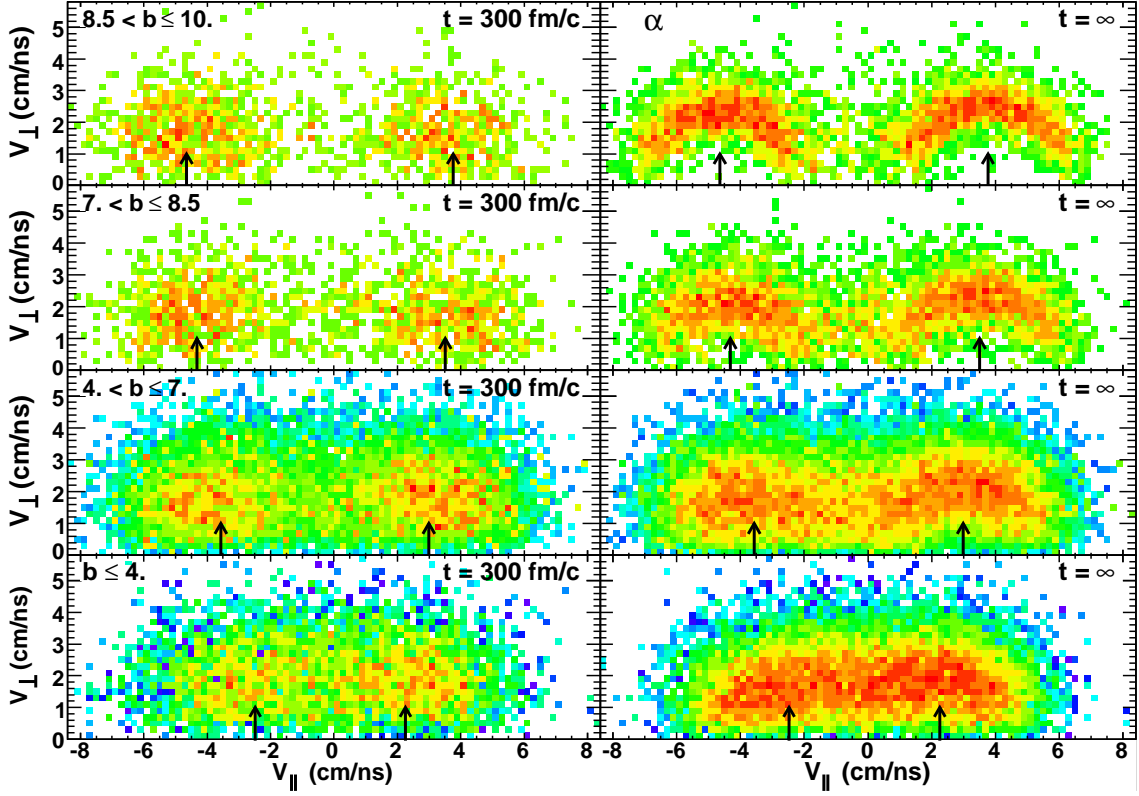}
\caption[]
{(Color online) 
Invariant cross-section for $\alpha$ particles in the COM frame.
The arrows indicate the average parallel velocity of the PLF$^*$ and TLF$^*$. 
The color scale indicates the yield on a logarithmic scale. In the left column
the vertical scale of the bottom panel is scaled by a factor of six as compared
to the upper panels. The right column is scaled by a factor of three with 
respect to the left column.}
\label{fig:Gal_alpha}
\end{figure*}

For peripheral collisions, it has been experimentally observed that 
the emission pattern of $\alpha$ particles emitted by the PLF$^*$ 
manifests an anisotropic distribution \cite{Hudan04}.
This observed anisotropy has been interpreted as the decay of
a PLF$^*$ (and TLF$^*$) initially deformed by the collision process. To 
investigate the extent to which such a physical picture is compatible with the 
AMD model, we have examined the invariant cross-section 
maps of $\alpha$ particles. 
Depicted in Fig.~\ref{fig:Gal_alpha} is the dependence of the invariant 
cross-section map for $\alpha$ particles on impact parameter 
both at t=300 fm/c and following Coulomb propagation to 
infinite separation and sequential decay. At t=300 fm/c (left column), 
for the most peripheral collisions the yield is peaked near
the average PLF$^*$ and TLF$^*$ velocities (indicated by arrows), though
slightly toward the center-of-mass velocity.
For these most peripheral collisions one observes that the primary $\alpha$
yield centered at mid-velocity is relatively small.
With decreasing impact parameter,
the primary alpha distributions associated with the PLF$^*$ and TLF$^*$ 
move closer in velocity and increasingly overlap. 

Following secondary decay, the Coulomb circles evident for $\alpha$ particles are even more striking than
those for IMFs. This observation is consistent with the large multiplicity of
$\alpha$ particles that originate from the de-excitation of the 
PLF$^*$ and TLF$^*$ as compared to the early dynamical stage.
The distinct emission pattern observed for 8.5$<$b$\le$10 fm is 
also observed for more 
central collisions although with increasing centrality 
the distinct nature of the 
semi-circles becomes less striking. The ridge of yield which is typically
interpreted as Coulomb barrier emission becomes broader and 
its center moves increasingly toward 
V$_{\parallel}$=0. These trends are consistent with the increased 
damping, excitation, and reduced size of the PLF$^*$ (TLF$^*$) with increasing 
centrality. For even the most peripheral collisions, the pattern 
evident in the right column of Fig.~\ref{fig:Gal_alpha} 
is clearly not isotropic, favoring backward emission. 
As the sequential decay of the PLF$^*$ following t=300 fm/c is taken 
to be that of an isolated spherical nucleus without considering 
the influence of the external Coulomb field of the target on its decay
\cite{Hudan04}, 
it does not contribute to the predicted anisotropy. 
Within the model calculation, the observed anisotropy 
has two possible origins: Coulomb focusing of the $\alpha$ 
particles present at t=300 fm/c and the $\alpha$ decay of 
IMFs which are emitted anisotropically.


\begin{figure}
\vspace*{4.0in}
\includegraphics{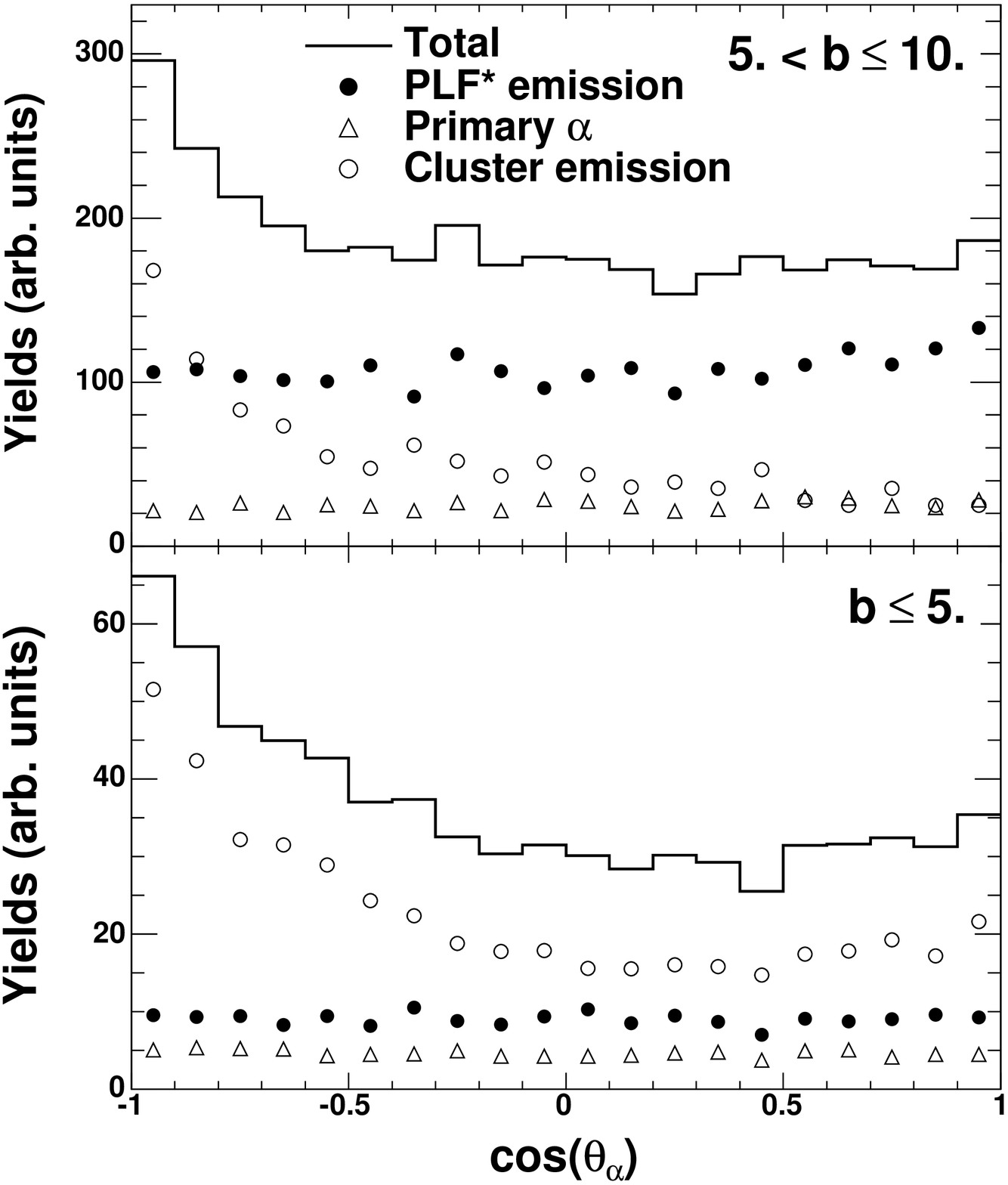}
\caption[]
{Angular distribution for $\alpha$ particles on the PLF$^*$ Coulomb ridge 
(V$_\alpha$$<$ 3.5 cm/ns). All angles are in the PLF$^*$ frame.}
\label{fig:Alpha_ridge}
\end{figure}

The anisotropic emission of $\alpha$ particles along the PLF$^*$ 
Coulomb ridge has recently been proposed to be 
related to the enhanced backward decay of the 
excited PLF$^*$ due to the nucleus-nucleus interaction \cite{Hudan04}. 
Displayed in Fig.~\ref{fig:Alpha_ridge} is the $\alpha$ particle yield 
along the Coulomb ridge
for ``peripheral'' collisions, 5$<$b$\le$10 fm, and ``central'' collisions, 
b$\le$5 fm.
Alpha particles were selected to be ``Coulomb barrier'' particles by 
restriction on their velocities, namely V$_\alpha$$<$ 3.5 cm/ns in the PLF$^*$ 
frame. In both cases shown, the total $\alpha$ 
particle yield (solid histogram) is not symmetric with
respect to emission transverse to the PLF$^*$ direction, 
namely cos($\theta_\alpha$)=0. Emission in the backward direction 
cos($\theta_\alpha$)$<$0 is enhanced with respect to the forward direction.
This enhancement is more pronounced for the central collisions.
For the peripheral collisions, the emission yield
for cos($\theta_\alpha$)= -1 is approximately 1.7 times the yield
emitted in the transverse direction. In contrast, the forward emission yield
cos($\theta_\alpha$)= +1 is approximately the same as the transverse yield.
Comparison of the integrated yield with -1$\le$cos($\theta_\alpha$)$<$0,
Y$_{backward}$($\alpha$), to
0$<$cos($\theta_\alpha$)$\le$+1,
Y$_{forward}$($\alpha$), reveals that backward emission is enhanced
by $\approx$19\% as compared to forward emission. 
For more central collisions (bottom panel), 
comparison of the integrated yield 
reveals that backward emission is enhanced
by $\approx$39\% as compared to forward emission.

We have investigated the origin of this backward enhancement, by examining the
possible sources of $\alpha$ particles. Alpha particles are ``tagged'' as 
being either
a) ``primary'', namely those originating at the time of 
cluster recognition (t=300 fm/c), 
b) PLF$^*$ alphas or c) cluster alphas i.e. those 
that result from 
the secondary decay of primary IMFs. 
As expected, PLF$^*$ 
emission is essentially isotropic. 
It is evident in
Fig.~\ref{fig:Alpha_ridge} that for both peripheral and central collisions, 
primary alphas (open triangles) on the PLF$^*$ Coulomb ridge are isotropic. 
Evidently the Coulomb focusing of primary $\alpha$ particles by the 
PLF$^*$ and TLF$^*$ does not contribute to the anisotropy observed in 
Fig.~\ref{fig:Gal_alpha}.
The large backward enhancement observed for the total $\alpha$ particle 
yield is associated with the $\alpha$ particles 
that originate from the secondary decay of primary IMFs. 
Hence, it is the anisotropy of the primary IMFs 
that is responsible for the anisotropy of $\alpha$ particles 
associated with Coulomb barrier energies. 
Quantitative comparison of the 
the various components reveals that for peripheral collisions 
the ratio Y$_{backward}$($\alpha$)/Y$_{forward}$($\alpha$) associated with the
decay of primary clusters is $\approx$2.2, while the same ratio for
PLF$^*$ emission or primary $\alpha$ emission is $\approx$0.9.
In the case of more central collisions, 
the ratio Y$_{backward}$($\alpha$)/Y$_{forward}$($\alpha$) associated with the
decay of primary clusters is $\approx$1.7. The yield ratio for 
PLF$^*$ emission and primary $\alpha$ emission is $\approx$1.0 
in this impact parameter interval.

The anisotropy observed for the decay of 
primary clusters is consistent with the emission pattern of IMFs 
as shown in Fig.~\ref{fig:Gal_IMF}. This feeding of 
$\alpha$ particles to the PLF$^*$ Coulomb ridge from IMF secondary decay 
is also consistent with the fact that IMFs are produced excited, 
even for the most peripheral collisions. 
In fact, the average excitation energy of the IMFs is relatively independent of 
the impact parameter. For the most peripheral collisions,
$\langle$E$^*$/A$\rangle$ of the IMFs is typically 2.5 to 3 MeV 
with the higher values associated with IMFs produced around 
the center-of-mass velocity.
With increasing centrality, $\langle$E$^*$/A$\rangle$ becomes independent of
the IMF velocity and reaches a typical value of 3 MeV. 
Such an excitation energy is in agreement with the excitation energy 
experimentally  deduced for IMFs produced in central 
collisions \cite{Hudan03}.  
Investigation of the width of the IMF excitation energy distribution 
reveals that it is large and 
approximately independent of the impact parameter.

In addition to the anisotropies predicted by the model, in reality  
the anisotropic emission 
pattern of $\alpha$ particles can have additional origins. 
Although the PLF$^*$ is clearly deformed for t$\le$300 fm/c (as shown in 
Fig.~\ref{fig:density}), 
the statistical decay of the PLF$^*$ (and TLF$^*$) is assumed to be isotropic.
However, if the collision dynamics preferentially ``prepares'' the system in 
a configuration that favors emission toward the center-of-mass, the observed
emission pattern will certainly be anisotropic. An example of such a favored 
configuration would be a di-nuclear configuration of the PLF$^*$ decaying into
an IMF and residue with the IMF preferentially oriented toward mid-rapidity.
If the di-nuclear configuration prepared lies outside the saddle point 
for such a system, then the excitation energy of the di-nuclear configuration
does not influence the decay probability
and the decay is clearly non-statistical. However, if the di-nuclear 
configuration lies inside the saddle point, excitation energy does 
influence the decay probability and the emission can be considered statistical.
In this case, explicit treatment of the deformation within a statistical 
framework is necessary \cite{Charity05}.
The observed anisotropy under such conditions will 
depend sensitively on the emission time
relative to the rotational period of the di-nuclear system. Of course such a 
schematic description of the binary decay of the PLF$^*$ could be extended to
ternary and quaternary decays.
It should also be noted that such short timescale emission 
when the nuclei are 
in proximity of each other and can also be influenced 
by tidal effects \cite{Charity01}.

\section{Summary and Conclusions}

Using the AMD model,  we have examined the dynamical phase of a 
heavy-ion collision at intermediate energy. 
We have investigated how observables such as the size (Z), velocity, 
and excitation of the reaction products evolve during the early stages of the 
collision. The de-excitation of the initial reaction products is 
calculated with a statistical decay code
and the survival of these initial observables is examined.

We have investigated how the characteristics of the two large
remnants in the reaction evolve with impact parameter.
Both the 
$\langle$Z$\rangle$ and $\langle$V$\rangle$ of the PLF$^*$ and TLF$^*$ 
decrease smoothly as centrality increases 
up to an impact parameter of $\approx$4 fm. 
As the centrality increases from the most peripheral collisions, 
the PLF$^*$'s velocity is increasingly damped 
from the projectile velocity. Concurrent with this damping, 
the width of the velocity distribution increases. 
Although the average velocity is largely unchanged 
by secondary decay, the width of the velocity distribution
is typically increased by 10-40 \%.
For smaller impact parameters, b$<$3-4 fm, the average atomic number and 
velocity of the two reaction partners are independent of impact parameter. 
Associated with these changes in the size and velocity
and size of the PLF$^*$ and TLF$^*$, one also observes that 
the $\langle$E$^*$/A$\rangle$ of the PLF$^*$ and TLF$^*$ increases 
as the impact parameter decreases
from an initial value of 0.7-1.1 MeV upto 4 MeV. 
The maximum excitation energy is attained for an impact
parameter of $\approx$6 fm. Smaller impact parameters do not result in
larger values of $\langle$E$^*$/A$\rangle$.
These observations suggest that the peripheral collisions on one side and 
the most central collisions on the other side correspond to different 
dynamics regime although simulated with the same ingredients.

Peripheral collisions, as may be expected, exhibit a binary nature 
with a strong memory of the entrance channel.
In such collisions, a transiently deformed PLF$^*$ and TLF$^*$ 
are recognizable as early as
$\approx$100 fm/c after the collision. 
The deformation of these reaction products persists for a considerable time, 
t$\ge$300 fm/c. 
In addition to the PLF$^*$ and TLF$^*$, nucleons, light charged particles,
and IMFs are also produced in the dynamical phase. The latter clusters,
are preferentially located between the PLF$^*$ and TLF$^*$.
The Z distribution of particles with  V$_\parallel$$>$0 strongly favors 
asymmetric splits. The population of symmetric splits increases for
mid-peripheral collisions reflecting an increase in the excitation energy 
of the PLF$^*$. The excitation energy of the PLF$^*$ is strongly correlated 
with its velocity damping.
Both the $\langle$E$^*$/A$\rangle$ and $\langle$V$_{PLF^*}$$\rangle$ 
in this impact parameter range 
are found to be independent of the cluster recognition time.
The correlation between the excitation energy of the PLF$^*$ 
and velocity damping is the same for the different cluster recognition times 
studied. The general insensitivity of $\langle$V$_{PLF^*}$$\rangle$ 
to cluster recognition time makes it a robust signal of the impact parameter.
The excitation energy of the PLF$^*$ is slightly correlated with 
the excitation energy of the TLF$^*$ for early cluster recognition times. 
Even the small particle emission that occurs on 
short timescale is sufficient to destroy 
this correlation by t=300 fm/c.

In contrast to peripheral collisions which exhibit a strong binary character, 
the most central collisions do not manifest as much memory of the 
entrance channel. 
Such collisions are no longer dominated by two large fragments, 
namely the PLF$^*$ and TLF$^*$. However, if we designate the largest fragment 
forward and backward of the center-of-mass as the  PLF$^*$ and TLF$^*$, 
their characteristics, $\langle$Z$\rangle$, $\langle$V$\rangle$ and 
$\langle$E$^*$/A$\rangle$, are largely unchanged as b decreases 
for b$\le$4 fm. 
Therefore, for the innermost $\approx$10 \% of the total cross-section,  
the maximum degree of excitation for such collisions is attained. 
This broad range of impact parameters associated with high excitation 
underscores the importance of considering the breakup of non-spherical 
geometries \cite{leFevre99}.
Moreover, for these small impact parameters, the 
quantitative characteristics of 
the PLF$^*$ and TLF$^*$, $\langle$Z$\rangle$ and $\langle$E$^*$/A$\rangle$, 
depend on the cluster recognition time.
For early cluster recognition time (t=150 fm/c) an average 
excitation energy of 6 MeV 
is reached, while a longer cluster recognition time (300 fm/c) results in 
$\langle$E$^*$/A$\rangle$ of  4 MeV. This 
decrease in $\langle$E$^*$/A$\rangle$ 
indicates a rapid de-excitation during the dynamical stage, 
suggesting significant nucleon and cluster emission on a 
short timescale.

Direct examination of the multiplicities of emitted particles reveals that
the IMF multiplicity increases smoothly with increasing centrality 
and saturates for an impact parameter of $\approx$3 fm. 
At t=300 fm/c, the average IMF multiplicity reaches unity 
for an impact parameter of $\approx$9 fm. 
At all impact parameters and for t$>$150 fm/c, the IMF emission rate 
decreases monotonically with increasing cluster recognition time.  
The velocity distribution of the produced IMFs at t=300 fm/c is bimodal 
and reveals preferential emission from the 
PLF$^*$ and TLF$^*$ 
towards the center-of-mass. 
The emission pattern of $\alpha$ particles at t=$\infty$ also exhibits a 
distinct preferential emission towards the center-of-mass.
This anisotropy however, is not due to the anisotropic emission of primary 
$\alpha$ particles or 
evaporation from the PLF$^*$ and TLF$^*$, 
but arises from the secondary decay of anisotropically emitted primary IMFs.  
The multiplicity of neutrons saturates for mid-central collisions, making 
neutron multiplicity a poor selector of central collisions. The geometric 
cross-section combined together with the saturation of excitation energy for 
mid-peripheral collisions may explain the observed persistence of 
binary collisions at intermediate energies even when the largest 
neutron multiplicities are selected \cite{Lott92}.

The large excitation energy reached in the collision 
leads to rapid particle emission on the dynamical timescale. However, the 
present treatment of  
the short timescale decay involves several simplifications. 
The role of deformation in the decay is neglected as are both Coulomb 
and nuclear proximity effects. In addition the excitation energy is 
calculated relative to clusters that are at the ground-state both in 
shape and density. These simplifications may have a non-negligible impact on 
the characteristics of the fragmenting system.
This rapid de-excitation emphasizes the need for 
a hybrid statistical-dynamical model that considers in a more realistic manner
the statistical decay of the transiently deformed nuclei 
from times as short as 100 fm/c.
Development of such a hybrid model would represent a new and potentially 
powerful tool in understanding the dynamics of intermediate 
energy heavy-ion collisions, as well as cluster formation on short timescales.

\begin{acknowledgments}

This work was supported by the
U.S. Department of Energy under DE-FG02-92ER40714 (IU)
and  in part by Shared University Research grants from 
IBM, Inc. to Indiana University. In addition, A. O. would like to thank the National 
Superconducting Cyclotron Laboratory at Michigan State University for the 
warm hospitality
extended to him during his long term stay.

\end{acknowledgments}

\bibliography{amd.bib} 

\end{document}